\documentclass[10pt,aps,prb,twocolumn,amsmath,amssymb,longbibliography]{revtex4-1}
\usepackage{amsmath}
\usepackage{amssymb}
\usepackage{xcolor}
\usepackage{graphicx}
\usepackage[caption=false]{subfig}
\usepackage{multirow}
\usepackage{array}
\usepackage{color}

\graphicspath{{graphs/}}

\newcommand{\bra}[1]{\ensuremath{\langle #1 \vert}}
\newcommand{\ket}[1]{\ensuremath{\vert #1  \rangle}}
\newcommand{\braket}[2]{\ensuremath{\langle  #1 \vert #2  \rangle}}
\newcommand{\expect}[3]{\ensuremath{\langle  #1 \vert #2 \vert #3 \rangle}}

\renewcommand\b[1]{\ensuremath{\mathbf{#1}}}
\newcommand\B[1]  {\ensuremath{\pmb #1}}
\newcommand\equ[2]{\begin{align}#2\label{#1}\end{align}}

\newcommand\etal{\textit{et al.}~}

\newcolumntype{P}[1]{>{\centering\arraybackslash}p{#1}} 

\date{\today}

\begin{document}

\author{James E. T. Smith}
\email{james.e.smith@colorado.edu}
\affiliation{Department of Chemistry and Biochemistry, University of Colorado Boulder, Boulder, CO}
\author{Bastien Mussard}
\affiliation{Department of Chemistry and Biochemistry, University of Colorado Boulder, Boulder, CO}
\author{Adam A. Holmes}
\affiliation{Department of Chemistry and Biochemistry, University of Colorado Boulder, Boulder, CO}
\author{Sandeep Sharma}
\email{sandeep.sharma@colorado.edu}
\affiliation{Department of Chemistry and Biochemistry, University of Colorado Boulder, Boulder, CO}

\title{Cheap and near exact CASSCF with large active spaces}

\begin{abstract}
	We use the recently-developed Heat-bath Configuration Interaction (HCI)
	algorithm as an efficient active-space solver to perform multi-configuration self-consistent field
	calculations (HCISCF) with large active spaces. We give a detailed derivation of the theory and show that difficulties associated with non-variationality of the HCI procedure can be overcome by making use of the Lagrangian formulation to calculate the HCI relaxed two body reduced density matrix. HCISCF is then used to study the electronic structure of butadiene, pentacene, and Fe-porphyrin. One of the most striking results of our work is that the converged active space orbitals obtained from HCISCF are relatively insensitive to the accuracy of the HCI calculation. This allows us to obtain nearly converged CASSCF energies with an estimated error of less than 1 mHa using the orbitals obtained from the HCISCF procedure in which the integral transformation is the dominant cost. For example, an HCISCF calculation on Fe-Porphyrin model complex with an active space of (44e, 44o) took only 412 seconds per iteration on a single node containing 28 cores, out of which 185 seconds were spent in the HCI calculation and the remaining 227 seconds were mainly used for integral transformation.  Finally, we also show that active-space orbitals can be optimized using HCISCF to substantially speed up the convergence of the HCI energy to the Full CI limit because HCI is not invariant to unitary transformations within the active space.
\end{abstract}

\maketitle

\section{Introduction}
Many molecular systems with interesting electronic structure exhibit strong correlation, and as a result they are not
well-described by the standard single-reference quantum chemical techniques, such as density functional theory~\cite{hohenberg1964inhomogeneous,kohn1965self,parr1994density}, M{\o}ller-Plesset~\cite{moller1934note} and coupled cluster theory~\cite{coester1958bound,vcivzek1966correlation,cizek1980coupled,purvis1982full}.
These highly multireference systems include transition metal complexes, excited states of conjugated organic molecules,
and systems far from equilibrium, such as those near the breaking or forming of a covalent bond.

A popular paradigm for generating a multi-determinant reference for such systems is to identify the ``active'' subset of the electrons and orbitals that are most
important for a correct qualitative description of the molecular physics, and assume that the remaining orbitals are either
always occupied (``core'' orbitals) or never occupied (``virtual'' orbitals). Once this partitioning of orbitals is chosen,
the active electrons can be fully correlated to obtain the complete active-space configuration interaction (CASCI) ground-state wavefunction and energy.
In addition to correlating the active-space electrons, the orbitals
can also be optimized such that the active orbitals contain the most important degrees of freedom.
The traditional heuristic method for finding these degrees of freedom is to minimize the CASCI energy,
resulting in the complete active-space self-consistent field~\cite{roos1980complete1,roos1980complete2,siegbahn1981complete,roos2007complete} (CASSCF) algorithm.

Due to the exponential scaling of an exact correlated calculation with system size,
the maximum active-space size of a CASSCF is about 16 electrons in 16 orbitals, although recent developments have made it possible to go up to 20 electrons in 20 orbitals on massively parallel machines~\cite{cas}.
Naturally, there is great interest in developing algorithms that can overcome this limit in a systematically improvable way.\textbf{}
There are many algorithms that can be used as approximate active-space solvers,
including the density matrix renormalization group\cite{White1992,White1993,Fano1998,White1999,sch05,legeza2015,sch11,Daul2001,Chan2002, Moritz2006,Zgid2008,Luo2010,Marti2010,Chan2011,Kurashige2011,sharmaspin, sha-nat,kura-nat,wouters2014,keller,yuki-review,takeshi15}
restricted~\cite{malmqvist1990restricted,celani2000multireference} and generalized~\cite{ma2011generalized} active space methods, reduced density matrix approaches~\cite{NakEhaNak-JCP-02,mazziotti2006quantum,Val-ACP-07},
various flavors of selected configuration interaction followed by perturbation theory\cite{Ivanic2001,Huron1973,Buenker1974,Evangelista1983,Harrison1991,Steiner1994,Wenzel1996,Neese2003,Abrams2005,Bytautas2009,Evangelista2014,Knowles2015,Schriber2016,Liu2016,Caffarel2016,yann2017} (SCI+PT), and stochastic and semistochastic methods such as Full CI quantum Monte Carlo~\cite{Booth2009,Cleland2010,Petruzielo2012,thomas2015} (FCIQMC).

Here, we use the recently-developed Heat-bath Configuration Interaction~\cite{HolTubUmr-JCTC-16,ShaHolUmr-JCTC-17} (HCI) algorithm, an efficient SCI+PT algorithm, as an approximate active-space solver, in order to perform efficient CASSCF-like calculations with large active spaces. We call the resulting algorithm Heat-bath Configuration Interaction Self-Consistent Field (HCISCF). Similar extensions have already been presented with other theories such as DMRG-SCF\cite{macasscf,Ghosh2008,zgidcasscf}, FCIQMC-SCF\cite{Manni2016,thomas15}.
However, unlike DMRG-SCF, formulating HCISCF is made more complicated by the fact that HCI in not a variational method since it performs second-order perturbation theory. This implies that HCI energy is not stationary with respect to the variations of its zeroth-order wavefunction parameters $\b{c}$, \textit{i.e.} $\frac{\partial E_{\mathrm{HCI}}}{\partial \b{c}} \neq 0$. Thus, the first-order change in energy due to change in orbital rotation parameters $\B{\kappa}$ is given by

\begin{align}
	\frac{d E_{\mathrm{HCI}}}{d \B{\kappa}} = \frac{\partial E_{\mathrm{HCI}}}{\partial \B{\kappa}} + \frac{\partial E_{\mathrm{HCI}}}{\partial \b{c}}\frac{d\b{c}}{d \B{\kappa}}\label{eq:dedk},
\end{align}
where the second term does not vanish, as it does in DMRG-SCF and CASSCF. Here, and in the rest of the article, bold-face letters represent vectors. Equation~\ref{eq:dedk} suggests that to calculate the gradient of the HCI energy with respect to $\B{\kappa}$, one has to calculate the derivatives of the wavefunction with respect to each of the $O(n)$ parameters in $\B{\kappa}$, where $n$ is the number of basis functions (here we have assumed that the number of active and core orbitals is much smaller than the number of basis fuctions). In this work, we overcome
this prohibitive expense by using the technique of Lagrange multipliers, which is also known as the $z-$vector method.
The Lagrange multiplier technique replaces calculating the derivatives of the wavefunction with respect to all the parameters, with calculating only a single set of Lagrange multipliers.
It is worth mentioning that similar orbital optimization over a perturbatively-corrected energy is performed in orbital-optimized M{\o}ller-Plesset perturbation theory (OO-MP2).\cite{Lochan2007,Neese2009} 

A considerable simplification in the HCISCF theory arises if only the HCI variational energy is minimized, rather than the sum of the variational plus perturbative correction. In this case, the second term of Equation~\ref{eq:dedk} is zero because the variational HCI energy is stationary with respect to the wavefunction parameters $\b{c}$. We have implemented such a theory as well and call it vHCISCF to distinguish it from the full HCISCF.

We apply this new algorithm to calculate the ground states for butadiene, the pentacene monomer, and the Fe(II)-Porphyrin complex, abbreviated as Fe(P), as well as the excited states for the latter two systems.
For pentacene, we calculate the energies of the ground ($^1A_\text{g}$) and lowest triplet ($^3B_\text{2u}$) states, and compare the gaps to experimental\cite{BurgosJ.PopeM.SwenbergC.E.Alfano1977,BiermannD.Schmidt1979,Biermann1980} and theoretical\cite{Hachmann2007,Dorando2007,Hajgato2009,Zimmerman2010,Kurashige2014a,Coto2015,Yang2016} results of other CAS-based approaches. Fe(P) is a challenging electronic-structure problem, in which theory and experiments disagree on the symmetry of the ground state wavefunction.
\cite{Kim1975,Goff1977a,Walker2013,Dolphin1976,Lang1978,Boyd1979,Mispelter1980,Straws1985,Choe1998,Choe1999,Pierloot2009,Hachmann2007,Radon2008,Hajgato2009,Vancoillie2011,Manni2016,Phung2016} Due to the challenging nature of this system, we investigate the effects of basis set and choice of active-space orbitals on the gap between the $^5A_\text{g}$ and $^3B_\text{2g}$ states.

The remainder of the paper is organized as follows.
In Section~\ref{theory}, we briefly review the HCI algorithm with the aim of presenting the equations that will be used in the formulation of the HCISCF and vHCISCF algorithms. In Section~\ref{hciscf}, we derive the working equations of the HCISCF and vHCISCF algorithms. In Section \ref{comp_details}, we discuss the implementation and practical aspects of running an HCISCF calculation.
Finally, in Section \ref{results_and_disc}, we report our HCISCF calculations for butadiene, pentacene, and Fe(P), and compare them to experiments and previous calculations on these systems.

\section{The HCI algorithm}
\label{theory}
Heat-bath Configuration Interaction (HCI), like other SCI+PT schemes, consists of two stages:
\begin{itemize}
	\item a \textbf{variational stage} in which a set of important determinants is iteratively selected and used to compute a variational wavefunction and energy, and
	\item a \textbf{perturbative stage} in which the second-order correction to the variational energy is computed using multi-reference Epstein-Nesbet perturbation theory.
\end{itemize}

\begin{figure*}[!htbp]
	\subfloat[\label{fig:hci_a}]{\includegraphics[width=0.99\columnwidth]{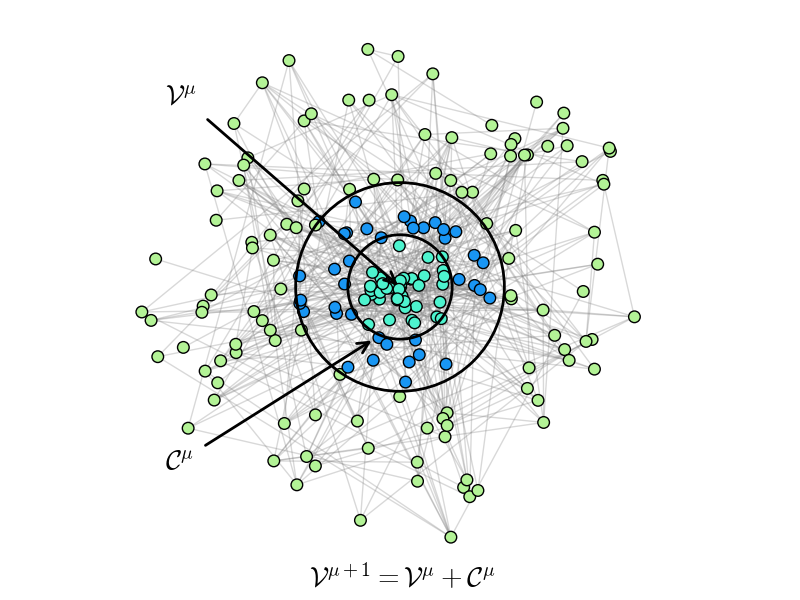}}
	\qquad
	\subfloat[\label{fig:hci_c}]{\includegraphics[width=0.99\columnwidth]{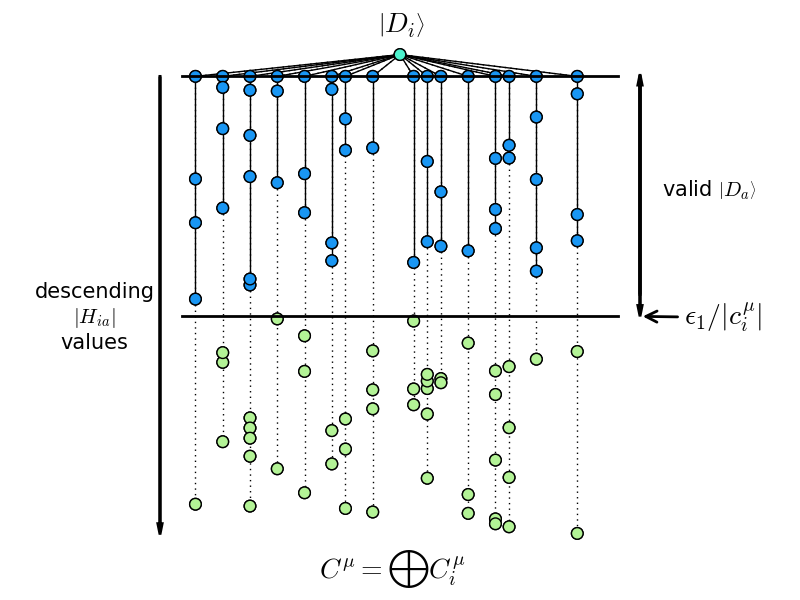}}
	\caption{Outline of the HCI scheme. The dots represent the determinants in a Hilbert space; the connections between them are randomized for the purpose of this sketch.
        Figure (a) shows that at a given iteration $\mu$ the current variational space is $\mathcal{V}^\mu$ (encompassing the cyan dots) and is augmented by the space of connected determinant $\mathcal{C}^\mu(\epsilon_1)$ (that includes the blue dots).
		This sequential aggregation of determinants is done iteratively until some designated convergence threshold is met.
		Figure (b) shows how HCI takes advantage of its importance function to add new determinants into $\mathcal{C}^\mu(\epsilon_1)$.
		The dotted lines represent possible excitations from every pair of occupied orbitals in a given cyan determinant $\ket{D_i}$ of a current $\mathcal{V}^\mu$.
        A determinant $\ket{D_a}$ is generated and included in $\mathcal{C}^\mu(\epsilon_1)$ if the \textit{magnitude} of $H_{ia}=\expect{D_i}{\hat{H}}{D_a}$ is greater than a certain $i$-dependent threshold, $\epsilon_1/\left|c_i^\mu\right|$.
		Hence, for all pairs of occupied orbitals in a given determinant (\textit{i.e.} for all dotted lines) the algorithm will only browse the sorted list of magnitudes up to that threshold and generate the corresponding blue determinants $\ket{D_a}$:
        no computational time is lost on generating determinants (here in green) that will not be included in $\mathcal{C}^\mu(\epsilon_1)$.}
	\label{fig:all}
\end{figure*}

\subsection{Variational Stage}

In the variational stage, a set of determinants $\mathcal{V}$ is iteratively generated
and their coefficients are variationally optimized to find a multi-determinantal
wavefunction and energy.
At a given iteration $\mu$, the current set of determinants in noted $\mathcal{V}^\mu$ and the current variational wavefunction is given by
\equ{eq:Psi0}{
    \ket{\Psi_0^\mu}=	\sum_{\ket{D_i}\in\mathcal{V}^\mu} c_i^\mu \ket{D_i},
}
where $\b{c}^\mu$ are the current CI coefficients.

The variational space is augmented (see Figure ~\ref{fig:hci_a}) as $\mathcal{V}^{\mu+1}=\mathcal{V}^\mu+\mathcal{C}^\mu(\epsilon_1)$ where $\epsilon_1$ is a user-defined parameter. The space of connected determinants $\mathcal{C}^\mu (\epsilon_1)$ is defined as
\begin{align}
	\mathcal{C}^\mu (\epsilon_1)=\big\{ \ket{D_a} ~\vert~ f^\mu(\ket{D_a})> \epsilon_1 \big\}, \label{eq:criteria}
\end{align}
where $f^\mu(\ket{D_a})$ is a non-negative ``importance'' function, which can vary between different SCI+PT schemes.

The HCI importance function is:
\begin{align}
	f_{\rm HCI}^\mu(\ket{D_a}) = \max_{ \ket{D_i} \in \mathcal{V}^\mu} \left|H_{ai} c^\mu_i\right|,
\end{align}
where $H_{ai}$ is the Hamiltonian matrix element between $\ket{D_a}$ and $\ket{D_i}$.
In other words, the space $\mathcal{C}^\mu(\epsilon_1)$ is composed of all determinants $\ket{D_a}$ for which $\left|H_{ai}c^\mu_i\right|>\epsilon_1$ is true for at least one determinant $\ket{D_i}$ in the current $\mathcal{V}^\mu$.

By contrast, the importance function of CIPSI and its variants is inspired by perturbation theory (hence the name ``CI by \emph{Perturbatively} Selecting Iteratively''), and is given by
\equ{eq:CIPSI}{
	f_{\rm CIPSI}^\mu(\ket{D_a}) =  \left| \frac{\sum_{\ket{D_i}\in \mathcal{V}^\mu}H_{ai}c^\mu_i }{E_0^\mu-E_a}\right|,
}
where $E_a$ is the energy of the determinant $\ket{D_a}$
and
$E_0^\mu$ is the current ground-state energy.

The advantage of the HCI importance function over the CIPSI one is twofold. First, the relevant information for the HCI importance function for a given reference (apart from its coefficient) is simply the magnitude of the connecting matrix element, $\left|H_{ai}\right|$.
Since the vast majority of excitations are double excitations, whose magnitudes are simple functions
of only the orbitals involved (and not the specific determinants involved), all the necessary
information can be sorted and stored prior to the run. As a result, the doubly-excited determinants meeting the
HCI criterion can be generated \emph{without generating lists of candidates first}, as would be
required for CIPSI (see Figure~\ref{fig:hci_c}). Second, the HCI importance function can be utilized in constant time, whereas the CIPSI importance function requires
performing a sum in the numerator (which may require communication across cores in a parallel run) and evaluating a diagonal element in the denominator.

After the addition of the determinants in $\mathcal{C}(\epsilon_1)$, the new ground-state energy and wavefunction within the new variational space are obtained at each iteration by the Davidson procedure, which amounts to minimizing the energy functional

\begin{align}
	E[\b{c}^\mu, \mathcal{E}_0^\mu; \mathcal{V}^\mu]=
    & \expect{ \Psi_0^\mu }{ \hat{H}_0 }{ \Psi_0^\mu }
    - \mathcal{E}_0^\mu (\braket{\Psi_0^\mu }{ \Psi_0^\mu}-1)
	,\label{eq:variational}
\end{align}
with respect to the coefficients $\b{c}^\mu$ of $\ket{\Psi_0^\mu}$ and to the Lagrange multiplier $\mathcal{E}_0^\mu$ ensuring that the new wavefunction remains normalized,
in the context of a space $\mathcal{V}^\mu$ which is fixed at this point.

\subsection{Perturbative stage}
Once a converged
variational space $\mathcal{V}$,
variational coefficients $\b{c}$,
variational wavefunction $\ket{\Psi_0}$,
and
corresponding variational energy $E_0$ have been obtained using the algorithm in the previous section, multireference perturbation theory can be performed to estimate the Full CI energy.
We use the Epstein-Nesbet partitioning of the Hamiltonian by defining the zeroth-order Hamiltonian
\begin{align}
    \hat{H}_0 = \sum_{\ket{D_i}, \ket{D_j} \in \mathcal{V}} H_{ij} \ket{D_i}\bra{D_j} + \sum_{\ket{D_a} \notin \mathcal{V}} H_{aa} \ket{D_a}\bra{D_a} \label{eq:h0}
,\end{align}
and perturbation $\hat{V} = \hat{H} - \hat{H}_0$. With this partitioning, the second-order perturbative energy correction to the variational energy is given by
\begin{align}
	E_2=
	\sum_{\ket{D_a}\in\mathcal{C}}
	\frac{1}
	{E_0-H_{aa}}\left(\sum_{\ket{D_i}\in\mathcal{V}} H_{ai}c_i\right)^2, \label{eq:pt}
\end{align}
where $\mathcal{C}$ denotes the set of determinants that are connected to at least one determinant in $\mathcal{V}$ by a non-zero Hamiltonian matrix element. The vast majority of the contributions in the double sum are negligibly small, and can be discarded without significant loss of accuracy.
HCI therefore approximates the perturbative energy correction as
\begin{align}
	E_2(\epsilon_2)=
    \sum_{\ket{D_a}\in\mathcal{C}(\epsilon_2)}
	\frac{1}
	{E_0-H_{aa}}\left(\sum_{\ket{D_i}\in\mathcal{V}}^{(\epsilon_2)} H_{ai}c_i\right)^2, \label{eq:pt_hci}
\end{align}
where the symbol $\sum^{(\epsilon_2)}$ denotes a ``screened sum'' in which terms smaller in magnitude than a user-defined parameter $\epsilon_2$ are discarded.
Hence, the important terms involve determinants in the set
$\mathcal{C}(\epsilon_2)$ that are connected to at least one determinant in $\mathcal{V}$ by a Hamiltonian matrix element larger in magnitude than $\epsilon_2/\left|c_i\right|$.
In the limit of $\epsilon_2 \rightarrow 0$, the exact perturbation correction is recovered.
In order to obtain a good approximation to the perturbative correction $\epsilon_2$ must be much smaller than $\epsilon_1$,
but a nearly exact approximation to the perturbative correction can be obtained at a much reduced cost by choosing a small but non-zero $\epsilon_2$ parameter.

The naive evaluation of Eq.~\ref{eq:pt_hci} requires one to simultaneously store the entire set of determinants $\mathcal{C}(\epsilon_2)$ in order
to combine contributions to the sum before they are squared. Even with the use of a non-zero parameter $\epsilon_2$, the number of determinants in $\mathcal{C}(\epsilon_2)$ can be extremely large, resulting in a memory bottleneck.

The second-order energy correction can alternatively be estimated stochastically with the same accuracy, circumventing the memory bottleneck at the cost of the intrusion of an unbiased stochastic error.
We refer the reader to Ref.~\citenum{ShaHolUmr-JCTC-17} for details on this unbiased sampling procedure, which uses a sampled subset of the determinants in the zeroth-order wavefunction.

The stochastic error can be reduced by implementing a \emph{semistochatic} algorithm, in which a deterministic perturbative calculation is performed using a large parameter $\epsilon_2^d$ to avoid the memory bottleneck, and the error is corrected stochastically using a tight parameter $\epsilon_2$, as follows (note the superscripts D and S for expressions that are evaluated deterministically or stochastically, respectively):
\begin{align}
    E_2(\epsilon_2) =  E_2^\text{D}(\epsilon_2^d) + \bigg[ E_2^\text{S}(\epsilon_2) -  E_2^\text{S}(\epsilon_2^d)\bigg]. \label{eq:semistoc}
\end{align}
The key point is that $E_2^\text{S}(\epsilon_2)$ and $E_2^\text{S}(\epsilon_2^d)$ are evaluated using the same set of sampled determinants; consequently, their errors are highly correlated and the difference between the two calculations (shown in the square brackets in Eq~\ref{eq:semistoc}) has a much reduced stochastic noise.

For the formulation of the self-consistent procedure, it is useful to point out that in place of Equation~\ref{eq:pt}, the second-order correction to the energy can equivalently be obtained by minimizing the Hylleraas functional
\begin{align}
    H[\b{d}; \b{c}, \mathcal{V}, \mathcal{C}(\epsilon_2)]
   &= \expect{\Psi_1^{\epsilon_2} } {\hat{H}_0-E_0 } {\Psi_1^{\epsilon_2}}
   +2\expect{\Psi_1^{\epsilon_2} } {\hat{V}       } {\Psi_0} \label{eq:hylleraas}
,\end{align}
with respect to the coefficients $\b{d}$ of the first-order wavefunction
\begin{align}
    \ket{\Psi_1^{\epsilon_2}}=\sum_{a \in \mathcal{C}(\epsilon_2)} d_a \ket{D_a}
,\end{align}
with $\b{c}$, $\mathcal{V}$ and $\mathcal{C}(\epsilon_2)$ held fixed.
At its minimum, the Hylleraas functional gives the optimal values of $\b{d}$, and the value of the functional is equal to the second-order correction evaluated using Eq~\ref{eq:pt}. This will be used in Section~\ref{hciscf} to derive formulas for HCISCF calculations.

It should be mentioned at this point that $E_2(\epsilon_2)$ is not a strict upper limit to $E_2$, since the introduction of the non-zero parameter $\epsilon_2$ not only truncates the size of the space $\mathcal{C}(\epsilon_2)$, but also changes the perturbation $\hat{V}$ by ignoring small matrix elements. The Hylleraas functional formulation of perturbation theory seems to show that merely truncating the size of $\mathcal{C}(\epsilon_2)$ would provide a variational upper bound to the second-order energy
\textit{i.e.} $E_2(\epsilon_2) \geq E_2$; however, because the perturbation $\hat{V}$ is simultaneously changed, the strict variationality is destroyed. Instead, given $\epsilon_2$ and the zeroth-order wavefunction ($\b{c}$ and $\mathcal{V}$), the inequality $H[\b{d}; \b{c}, \mathcal{V}, \mathcal{C}(\epsilon_2)] \geq E_2(\epsilon_2)$ holds, where the equality holds at its minimum.

\section{HCI self-consistent field }
\label{hciscf}

The HCISCF algorithm presented here is designed as a CASSCF-like procedure, in which the CASCI is replaced by an HCI calculation in the active space. However, an important distinction between HCISCF and CASSCF is that unlike CASCI, the unconverged active-space HCI energy is in general not invariant to active-active rotations. For instance, using natural orbitals as opposed to canonical orbitals can result in a drastically improved convergence of HCI to the CASCI or Full CI limit. This flexibility will be utilized to optimize the active-space orbitals in order to accelerate convergence.

Since HCI uses both a variational step and second-order perturbation theory, we have the choice of optimizing the total HCI energy or just the variational HCI energy with respect to the orbital coefficients. In this section we describe both these procedures and call them HCISCF and vHCISCF respectively.

\subsection{HCISCF}

The HCI energy is given by the sum of the zeroth-order and second-order energies calculated by minimizing respectively the energy functional in Eq~\ref{eq:variational} with respect to $\b{c}$ and $\mathcal{E}_0^\mu$, and the Hylleraas functional in Eq~\ref{eq:hylleraas} with respect to $\b{c}, \mathcal{E}_0$, and $\b{d}$.
Formally, the minimization of these functionals is performed by setting to zero their partial
derivatives with respect to the parameters $\b{c}, \mathcal{E}_0$ and $\b{d}$.
Thus the most natural way of deriving the HCISCF procedure is by setting to zero the derivative of the HCI functional with respect to the parameter $\B{\kappa}$ of the orbital coefficients.
The HCI functional reads:
\begin{align}
	E_\text{HCI}[\B{\kappa}, \b{c}, \mathcal{E}_0, \b{d}]=E[\B{\kappa}, \b{c}, \mathcal{E}_0] + H[\B{\kappa}, \b{c}, \b{d}]
,\end{align}
where the dependence on the parameters $\B{\kappa}$ is shown and the dependence on $\mathcal{V}$ and $\mathcal{C}$ is dropped.
Its derivative with respect to $\B{\kappa}$ reads:
\begin{align}
	\frac{d E_{\mathrm{HCI}}}{d \B{\kappa}}
	  & = \frac{\partial E_{\mathrm{HCI}}}{\partial \B{\kappa}}
	+ \frac{\partial E_{\mathrm{HCI}}}{\partial \b{c}}\frac{d\b{c}}{d \B{\kappa}}
	\nonumber\\&\quad
	+ \frac{\partial E_{\mathrm{HCI}}}{\partial \mathcal{E}_0}\frac{d\mathcal{E}_0}{d \B{\kappa}}
	+ \frac{\partial E_{\mathrm{HCI}}}{\partial \b{d}}\frac{d\b{d}}{d \B{\kappa}}
	\label{eq:dedk2}.
\end{align}
As mentioned in the introduction, this formulation requires the calculation of the derivatives of the HCI parameters with respect to $\B{\kappa}$, which makes it an impractical approach.

Instead, we introduce the Lagrangian
\begin{widetext}
	\begin{align}
		\mathcal{L}[\B{\kappa}, \b{c}, \mathcal{E}_0, \b{d}, \B{\lambda}_\text{c}, \B{\lambda}_\text{d}]
		= E[\B{\kappa}, \b{c}, \mathcal{E}_0]
		+ H[\B{\kappa}, \b{c}, \b{d}]
		+ \B{\lambda}_\text{c}^\dag \frac{\partial E}{\partial \b{c}^\dag }
		+ \B{\lambda}_\text{d}^\dag \frac{\partial H}{\partial \b{d}^\dag },
	\end{align}
\end{widetext}
which is a function of the variables $\B{\kappa}$, $\b{c}$, and $\b{d}$, and the set of Lagrange multipliers $\mathcal{E}_0$, $\B{\lambda}_\text{c}$, and $\B{\lambda}_\text{d}$. If the partial derivatives of the Lagrangian with respect to the Lagrange multipliers are set to zero, one obtains the governing equations of HCI from which the optimal values of $\b{c}$ and $\b{d}$ can be recovered.
Note that at these optimal values of $\b{c}$ and $\b{d}$, the Lagrangian is by construction equal to the HCI energy functional.
Furthermore, the Lagrange multipliers $\B{\lambda}_\text{c}$ and $\B{\lambda}_\text{d}$ can be used to impose the stationarity of the Lagrangian with respect to $\b{c}$ and $\b{d}$. Thus, the Lagrangian can be made stationary with respect to all its parameters, allowing the calculation of the derivative of the HCI energy with respect to $\B{\kappa}$ as

\begin{align}
\frac{d E_{\mathrm{HCI}}}{d \B{\kappa} } &= \frac{d \mathcal{L}}{d \B{\kappa} } = \frac{\partial \mathcal{L}}{\partial \B{\kappa} }
,
\end{align}
leading to the desired minimization of the HCI energy with respect to $\B{\kappa}$ under the constraint that $\frac{\partial E}{\partial \b{c} }=0$ and $\frac{\partial H}{\partial \b{d} }=0$.

Let us first derive the equations to obtain the Lagrange multipliers:
\begin{align}
	\frac{\partial \mathcal{L}}{\partial \b{d}} & = \frac{\partial H}{\partial \b{d} } + \B{\lambda}_\text{d}^\dag \frac{\partial^2 H}{\partial \b{d} \partial\b{d}^\dag }                                                                                                                        & =0 \label{eq:lagd} \\
	\frac{\partial \mathcal{L}}{\partial \b{c}} & = \frac{\partial E}{\partial \b{c} } + \frac{\partial H}{\partial \b{c} } + \B{\lambda}_\text{c}^\dag \frac{\partial^2 E}{\partial \b{c} \partial \b{c}^\dag }  + \B{\lambda}_\text{d} \frac{\partial^2 H}{ \partial \b{c}\partial \b{d}^\dag } & =0 \label{eq:lagc}
\end{align}
Here, we have assumed that all parameters are real numbers, although extension to complex numbers is straightforward without any additional complications. From Equation~\ref{eq:lagd}, we can infer that $ \B{\lambda}_\text{d}=0$ because  $\frac{\partial H}{\partial \b{d} }=0$. Equation~\ref{eq:lagc} simplifies to
\begin{align}
	\frac{\partial H}{\partial \b{c} } + \B{\lambda}_\text{c}^\dag \frac{\partial^2 E}{\partial \b{c}\partial \b{c}^\dag } & =0
,\end{align}
because $ \B{\lambda}_\text{d}=0$  and $\frac{\partial E}{\partial \b{c} }=0$. This equation can be solved to evaluate $\B{\lambda}_\text{c}$ (in the analytic gradient theory, this corresponds to the $z-$vector equation).

Finally, the gradient of the HCI energy with respect to \B{\kappa} is given by
\begin{widetext}
\begin{align}
    \frac{\partial \mathcal{L}}{\partial \B{\kappa} }
    &= \b{c}^\dag\frac{\partial H_0}{\partial \B{\kappa} }\b{c}
    +  \b{d}^\dag\frac{\partial H_0}{\partial \B{\kappa} }\b{d}
    + 2\b{d}^\dag\frac{\partial V}{\partial \B{\kappa} }\b{c}
    + 2\B{\lambda}_\text{c}^\dag\frac{\partial H_0}{\partial \B{\kappa} }\b{c} \nonumber\\
    &= \sum_{ijkl}\frac{\partial H_{0,ijkl}}{\partial \B{\kappa} } \Gamma_{ijkl}^{\text{c},\text{c}}
    +  \frac{\partial H_{0,ijkl}}{\partial \B{\kappa} } \Gamma_{ijkl}^{\text{d},\text{d}}
    + 2\frac{\partial V_{ijkl}}{\partial \B{\kappa} } \Gamma_{ijkl}^{\text{d},\text{c}}
    + 2\frac{\partial H_{0,ijkl}}{\partial \B{\kappa} } \Gamma_{ijkl}^{\lambda_\text{c},\text{c}}  \label{eq:lambda},
\end{align}
\end{widetext}
where the partial derivatives of the two-body integrals ($H_{0,ijkl}$ and $V_{ijkl}$) with respect to $\B{\kappa}$ are contracted with the transition two-body reduced density matrices ($\B{\Gamma}$; see Computational Details)
between two states shown as superscript (the one-body terms are not shown here to avoid proliferation of terms). A careful look at the equation reveals that the first term is the reduced density matrix of the variational wavefunction and the second two terms are the unrelaxed reduced density matrices of the perturbative correction, while the last term
arises due to the change in the second-order energy with the relaxation of the zeroth-order wavefunction as the orbitals are optimized. The partial derivatives of the two-body integrals with respect to $\B{\kappa}$ are calculated using the usual techniques, which are described in detail in Ref.\citenum{Sun}.

\begin{figure}[h]
	\centering
	\caption{\label{single_iter_scheme}
		The basic scheme for the HCISCF procedure is outlined here.
		The HCISCF module of PySCF is used to interface the Dice program with PySCF. Each iteration consists of a single Dice run, which returns the energy and the reduced density matrices to PySCF. These are in turn used to update the orbitals and the active space two electron integrals, which are passed on to Dice.
	}
	\includegraphics[scale=1.0,trim={2.00cm 1cm 2.0cm 1cm},clip,width=\linewidth]{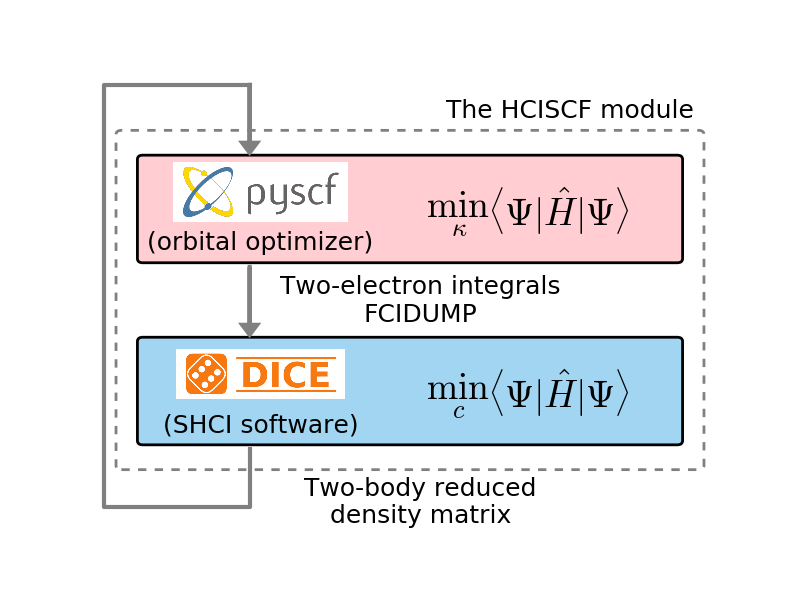}
\end{figure}

\subsection{$\mathrm{v}$HCISCF}
\label{vhciscf}
When only the variational energy $E_{\rm HCI}$ is optimized instead of the total HCI energy, one can calculate the gradient with respect to \B{\kappa} as
\begin{align}
    \frac{d E_\text{HCI}}{d \B{\kappa}} &=  \frac{\partial E}{\partial \B{\kappa}} + \frac{\partial E}{\partial \b{c}}\frac{d\b{c}}{d \B{\kappa}}\nonumber\\
&=\frac{\partial E}{\partial \B{\kappa}}\nonumber\\
&= \sum_{ijkl}\frac{\partial H_{0,ijkl}}{\partial \B{\kappa} } \Gamma_{ijkl}^{\text{c},\text{c}}\label{eq:vdedk2}.
\end{align}
Thus, to calculate the energy gradient, one does not need to evaluate the Lagrange multipliers, and the simple variational two-body reduced density matrix $\Gamma_{ijkl}^{\text{c},\text{c}}$ is sufficient.

\section{Computational Details}
\label{comp_details}

We briefly describe the calculation of the reduced density matrices encountered in Eq.~(\ref{eq:lambda}). The density matrices $\Gamma_{ijkl}^{\text{c},\text{c}}$ and $\Gamma_{ijkl}^{\lambda_\text{c},\text{c}}$ are strictly limited to states that contain determinants from the variational space $\mathcal{V}$. During the variational calculation, all single and double excitations between pairs of determinants in the variational wavefunction
are generated in order to evaluate the Hamiltonian. With this data available, the reduced density matrices can be calculated in a straightforward manner by looping over such connections and accumulating the contributions. The reduced density matrix $\Gamma_{ijkl}^{\text{d},\text{d}}$ contains the first-order wavefunction as both the bra and the ket; however, because $\hat{H}_0$ only contains the diagonal elements in the space of the connections $\mathcal{C}(\epsilon_2)$, the reduced density matrix
simply reads:

\begin{align}
    \Gamma_{ijkl}^{\text{d},\text{d}} = \sum_{\ket{D_a}\in\mathcal{C}(\epsilon_2)} d_a^2 \expect{D_a}{a_i^\dag a_j^\dag a_k a_l}{D_a}.
\end{align}
Finally, for the evaluation of $\Gamma_{ijkl}^{\text{d},\text{c}}$, all the determinants $\ket{D_i}$ in $\mathcal{V}$ that are connected by a Hamiltonian matrix element $\left|H_{ia}\right| > \epsilon_2$ to a determinant $\ket{D_a}$  are stored in a list. These connections are used at the end of the perturbative calculations to evaluate $\Gamma_{ijkl}^{\text{d},\text{c}}$ as
\begin{align}
    \Gamma_{ijkl}^{\text{d},\text{c}} = \sum_{\substack{\ket{D_i}\in\mathcal{V}\\\ket{D_a}\in\mathcal{C}(\epsilon_2)\\\left|H_{ia}\right| > \epsilon_2}} c_i d_a\expect{D_i}{a_i^\dag a_j^\dag a_k a_l}{D_a}.
\end{align}

Thus out of all the different reduced density matrices only the evaluation of $\Gamma_{ijkl}^{\text{d},\text{c}}$ adds a non-trivial memory cost over the HCI calculation, because not only do we have to store all the determinants $\ket{D_a}$ in the connected space $\mathcal{C}(\epsilon_2)$, but for each of these determinants we also have to store a list of variational determinants that are connected to them. This memory bottleneck can again be overcome with the use of semistochastic perturbation
theory, but in the paper we have not done so and have limited ourselves to using a value of $\epsilon_2$ for which the deterministic calculations can be performed. Although using the stochastic perturbation theory poses no challenge, we will see in the results section that the optimized orbitals converge relatively rapidly even when only a vHCISCF with a loose $\epsilon_1$ is performed.

All calculations in this work were performed using our HCISCF module in the PySCF software package, which interfaces PySCF with the Dice program. Dice is used to calculate the HCI energy and reduced density matrices. At each iteration of the HCISCF procedure, PySCF updates the orbital coefficients by calculating the energy gradient using the reduced density matrices obtained from the previous iteration. The updated orbitals are used to calculate the active-space Hamiltonian with which Dice
calculates the HCI energy and reduced density matrices. This procedure is illustrated in Fig \ref{single_iter_scheme} and is carried out until convergence. After convergence of the orbitals and CI coefficients, a final HCI calculation is performed where we use a smaller $\epsilon_1$ and $\epsilon_2$ to obtain near full configuration interaction (FCI) energy in the optimized active space.

\section{Results and discussion}
\label{results_and_disc}
We perform benchmark calculations on three different systems: butadiene, pentacene, and Fe-porphyrin. In addition to getting system-specific information, these calculations are meant to provide heuristics for running HCISCF calculations. We will investigate the following aspects:
\begin{itemize}
\item How tightly do we need to converge the HCI energies during the HCISCF procedure to obtain CASSCF-quality active-space orbitals?
\item Do vHCISCF and HCISCF converge to similar or substantially different active-space orbitals?
\item How much energy relaxation can be obtained just by optimizing the active-space orbitals, while keeping the active space itself fixed? Such a calculation where only the active-active orbital rotations are allowed during SCF will be called aHCISCF.
\end{itemize}
In the result section we use the acronyms vHCI and SHCI respectively to indicate the energies of the variational step and calculations where semistochastic perturbation theory was used.
All SHCI calculations performed here have stochastic noise of less than 0.05 mHa.

\subsection{Butadiene}
The HCI calculations on butadiene were performed with the same ANO-L-pVDZ basis set and geometry as the one used in Ref.~\citenum{Olivares-Amaya2015}. All electrons except the 1s orbitals were fully correlated to give an active space of (82o, 22e).

In this section, calculations were done using either Hartree-Fock canonical orbitals or optimized orbitals obtained from an aHCISCF calculation (where only the active-active rotations are allowed) with a relatively loose $\epsilon_1 = 3\times 10^{-4}$ Ha.
HCISCF energies typically are quadratically convergent because PySCF is able to perform pseudo-second-order optimization by estimating the Hessian, but the rate of convergence of aHCISCF calculations become substantially worse because of the strong coupling between the CI coefficients and orbital rotation parameters.
The cost of performing the orbital optimization is, however, more than made up for by the improved convergence of the HCI energies with the optimized orbitals.
Table~\ref{butadiene} shows that with approximately the same number of determinants in the variational space, the vHCI energy is more than 24 mHa lower when optimized active space orbitals are used as opposed to the Hartree-Fock canonical orbitals.
The effect is smaller for the full HCI energy, where the relaxation is only of 1.4 mHa, but one can observe that the rate of convergence of the full HCI energies with the number of determinants in the variational space still substantially improves when optimized orbitals are used.
As a result, the HCI energies calculated using the optimized orbitals can be accurately extrapolated to the FCI limit.

\begin{figure}
\begin{center}
\includegraphics[width=\linewidth]{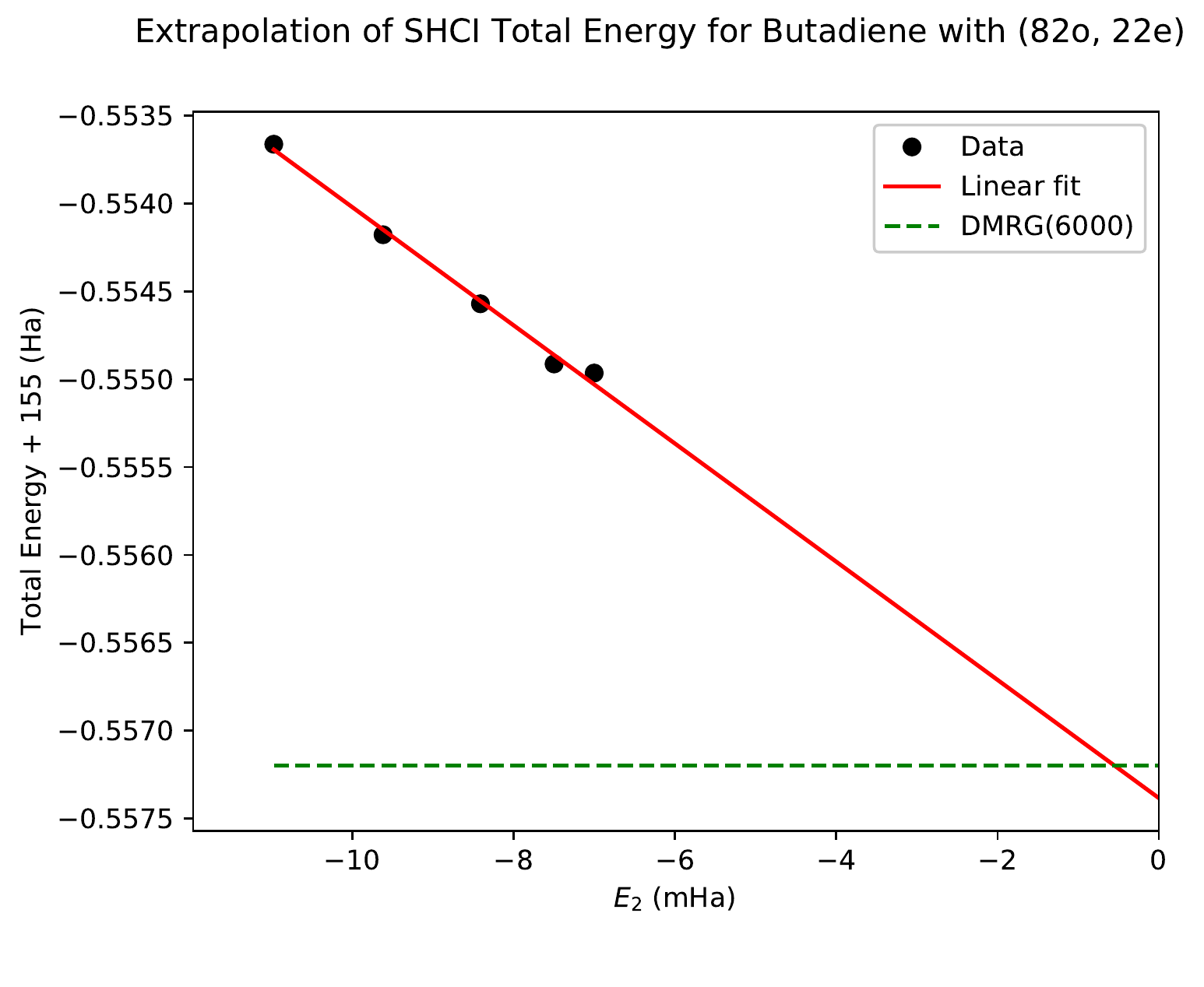}
\end{center}
\caption{Extrapolation of the SHCI total energy to the FCI limit for
the ground state of butadiene with an (82o, 22e) active space, using optimized aHCISCF orbitals. The dotted green line shows the DMRG energy calculated with a bond dimension of M=6000 and is believed to be converged to better than 1 mHa accuracy.
}\label{extrap}
\end{figure}

As in a previous paper~\cite{2017arXiv170803456H}, we extrapolate the HCI energy to the Full CI limit by extrapolating to where the perturbative correction goes to zero. In this work, we do this by first performing a single HCI calculation in which the smallest $\epsilon_1$ allowed by the available computational resources is used, which for butadiene is approximately $10^{-5}$ Ha. Some subsequent single step HCI calculations are performed in which a fraction of the least important determinants are discarded (in this work: 0.33). After a few such iterations, we plot the total HCI energy versus the PT correction and perform a linear extrapolation to $E_2\rightarrow 0$, as shown in Figure~\ref{extrap}. This procedure was inspired by the one used in DMRG calculations, where a large $M$ (number of retained renormalized states) calculation is performed, a few subsequent sweeps are carried out with progressively smaller values of $M$ and the energies and discarded weights obtained from these calculations are used to perform a linear extrapolation to a zero discarded weight.

\begin{table*}
 \caption{\label{butadiene} Results of HCI calculations on butadiene with an active space of (82o, 22e) with the ANO-L-pVDZ basis set. Two sets of HCI calculations were performed: calculations using the canonical Hartree-Fock orbitals (``Canonical'') and calculations using the aHCISCF optimized orbitals (``Optimized''). $N_\text{var}$, vHCI and SHCI respectively show the number of determinants included in the variational space, the variational energy (+155.0 Ha) and the final SHCI energy (+155.0 Ha) with stochastic error bar of 0.04 mHa.
The results of the calculations using the optimized orbitals were used to extrapolate the SHCI energy (see text for more details). The extrapolated SHCI energy is shown along with results obtained using other methods.}
 \renewcommand{\arraystretch}{1.2}
 \begin{tabular}{cP{17mm}P{17mm}P{17mm}
                 cP{17mm}P{17mm}P{17mm}}
 \hline
 \hline
 \multirow{2}{*}{$\epsilon_1$ (Ha)}&\multicolumn{3}{c}{Canonical}&&\multicolumn{3}{c}{Optimized}\\
  \cline{2-4}\cline{6-8}
  &$N_\text{var}$&vHCI&SHCI &
  &$N_\text{var}$&vHCI&SHCI \\
  \hline
  $3\times 10^{-5}$&$2.3\times 10^7$&-0.5195&-0.5526(1)&&$1.1\times 10^7$&-0.5411&-0.5534(1)\\
  $2\times 10^{-5}$&$4.8\times 10^7$&-0.5273&-0.5527(1)&&$2.1\times 10^7$&-0.5441&-0.5540(1)\\
  $1\times 10^{-5}$&-&-&-&&$5.9\times 10^7$&-0.5481&-0.5550(1)\\

\hline
&&&\multicolumn{3}{l}{\hspace*{7mm}SHCI($\epsilon_1 \rightarrow 0$)}&\multicolumn{2}{c}{-0.5574(8)}\\
&&&\multicolumn{3}{l}{\hspace*{7mm}CCSD(T)}                         &\multicolumn{2}{c}{-0.5550\phantom{(3)}}\\
&&&\multicolumn{3}{l}{\hspace*{7mm}CCSDT}                           &\multicolumn{2}{c}{-0.5560\phantom{(3)}}\\
&&&\multicolumn{3}{l}{\hspace*{7mm}DMRG(M=6000)}                    &\multicolumn{2}{c}{-0.5572\phantom{(3)}}\\
\hline
\hline
\end{tabular}
\end{table*}

\subsection{Pentacene}
\label{5cene_results}
Linear acenes, such as tetracene and pentacene, are promising candidates for singlet fission application because the gap to the lowest lying triplet state (T0) is roughly half of that of the first singlet excited state (S1). They also show great promise as organic semiconductors due to their large carrier mobility and low production cost. With recent applications to light emitting diodes, photovoltaic cells, and field effect transistors, these systems have been the subject of many theoretical and experimental studies.\cite{BurgosJ.PopeM.SwenbergC.E.Alfano1977,BiermannD.Schmidt1979,Biermann1980, Hachmann2007,Dorando2007,Hajgato2009,Zimmerman2010,Kurashige2014a,Coto2015,Yang2016}
The ground and excited state of acenes become progressively more multireference as their chain length is increased, and multireference methods are necessary to obtain an accurate description of their gap.

\begin{table}[h]
\caption{\label{eps1_and_rdm_comp}
  vHCISCF and HCISCF calculations performed on pentacene using various thresholds of $\epsilon_1$. The PT calculation in HCISCF was performed with $\epsilon_2=10^{-5}$ Ha. We also report extrapolated SHCI energies (see text) using the optimized active space orbitals obtained from the SCF calculations. It is interesting to note that although the HCISCF energies are far from converged, the SHCI energies agreement to within 0.1 mHa indicates that the active space orbitals are most likely converged.
}

\renewcommand{\arraystretch}{1.2}
\begin{tabular}{ P{0.13\columnwidth}P{0.1\columnwidth} P{0.2\columnwidth} P{0.25\columnwidth} P{0.2\columnwidth} }
\hline\hline
              &&  $\epsilon_1$ (Ha)         &  $E_\text{HCISCF}$ (Ha)  &  $E_\text{SHCI}$ (Ha)\\
    \hline
    vHCISCF   &&  $8.5\times 10^{-5}$  &  -841.5936          &  -841.6174 \\
    vHCISCF   &&  $5.0\times 10^{-5}$  &  -841.6005          &  -841.6175 \\
     HCISCF   &&  $8.5\times 10^{-5}$  &  -841.6021          &  -841.6173 \\
\hline\hline
\end{tabular}
\renewcommand{\arraystretch}{1}
\end{table}

We calculate the energies of the pentacene $^1A_\text{g}$ ground state and $^3B_\text{2u}$ lowest triplet state\cite{Hajgato2009,Zimmerman2010,Kurashige2014a,Yang2016} in the cc-pVDZ basis set on the optimized singlet and triplet geometries.
This allows us to calculate both vertical and relaxed (well-to-well) $^1A_\text{g} \rightarrow {}^3B_\text{2u}$ excitation energies. The active space chosen for all calculations is composed of the 11 $\pi$ and 11 $\pi^*$ orbitals.

We begin by assessing the effect of the accuracy of the HCI calculation on the active-space orbitals obtained after an HCISCF calculation. In Table~\ref{eps1_and_rdm_comp}, we show vHCISCF and HCISCF energies obtained using different $\epsilon_1$ cutoffs. We also show the energies obtained by performing a final ``tight'' HCI calculation on the three different optimized active space orbitals. It is interesting to note that even though the HCISCF energies are clearly un-converged, the final HCI energies agree to within 0.1 mHa in these three cases. This indicates that the optimized active-space orbitals obtained from the three SCF calculations are virtually identical.

Table~\ref{5cene_data} shows our gap results for pentacene. The HCI vertical gap for the singlet geometry are slightly higher than that of Kurashige \etal\cite{Kurashige2014a}; however, the difference of 1.5 kcal/mol can be due to the different basis set used, or to small differences in the choice of initial orbitals and subsequent convergence of the DMRG-SCF and HCISCF calculations.
Our vertical excitation energy calculated using the triplet geometry agrees reasonably well with the experimental excitation energy of 19.8 kcal/mol reported in Ref.~\citenum{BurgosJ.PopeM.SwenbergC.E.Alfano1977}
(a direct comparison with the Kurashige \etal paper is not pertinent, as they only report CASPT2 calculations).
However, it is worth noting that this energy is significantly different from both the vertical excitation energies calculated using the singlet geometry and the well-to-well excitation energies.

\begin{table*}[!htbp]
  \caption{\label{5cene_data}
  Pentacene ground (singlet) state and lowest triplet state energies
  calculated at the singlet and triplet geometries reported by Kurashige \etal\cite{Kurashige2014a}.
  An initial vHCISCF calculation was performed with an $\epsilon_1$ of 8.5$\times 10^{-5}$ Ha. The optimized active space obtained after this calculation was used to perform more accurate HCI calculations which were extrapolated to obtain near FCI energies with an estimated error shown in the table. $E_\text{ex}$ and $E_\text{ref}$ are respectively the vertical excitation calculated in this work and by Kurashige et al~\cite{Kurashige2014a}. $T_\text{OO}$ and $T_\text{CI}$, are the wall time in seconds needed for the HCI calculation and orbital optimization step during a single vHCISCF iteration, using a single node with two 14-core 2.4 GHz Intel$^\circledR$ Broadwell processors and a combined memory of 128 GB RAM.}

  \renewcommand{\arraystretch}{1.2}
  \begin{tabular}{ P{0.1\columnwidth} P{0.2\columnwidth} P{0.25\columnwidth} P{0.2\columnwidth} P{0.2\columnwidth} P{0.15\columnwidth} c }
\hline\hline
Sym.	&	$E_\text{vHCISCF}$	&	$E_\text{SHCI}$	&	$E_\text{ex}$	&	$E_\text{ref}$	&	$T_\text{OO}$	&	$T_\text{CI}$	\\
 & (Ha) & (Ha) & (kcal/mol) & (kcal/mol) & (sec) & (sec) \\
\hline
\multicolumn{7}{c}{}\\
\multicolumn{7}{c}{Singlet Geometry}\\
$^1A_\text{g}$	&	-841.5936	&	-841.6174(6)	&	\multirow{2}{*}{28.5}	&	\multirow{2}{*}{27.0}	&	50	&	33	\\
$^3B_\text{2u}$	&	-841.5457	&	-841.5720(8)	&		&		&	70	&	24	\\

\multicolumn{7}{c}{}\\
\multicolumn{7}{c}{Triplet Geometry}\\
$^1A_\text{g}$	&	-841.5823	&	-841.6050(7)	&	\multirow{2}{*}{18.6}	&	\multirow{2}{*}{-}	&	57	&	26	\\
$^3B_\text{2u}$	&	-841.5556	&	-841.5751(9)	&		&		&	57	&	31	\\
\hline\hline
  \end{tabular}
    \renewcommand{\arraystretch}{1}
\end{table*}

\subsection{Iron (II) Porphyrin}
\label{feP_results}

Fe(II) porphyrin (Fe(P)) are the active centers of several important biological proteins such as hemoglobin, myoglobin and catalase.
Experimental work suggests that the ground state of Fe(P) is a triplet state belonging to either the $^3A_\text{2g}$ or the $^3E_\text{g}$ irreducible representation of the $D_\text{4h}$ point group \cite{Kim1975,Goff1977a,Walker2013,Dolphin1976,Lang1978,Boyd1979,Mispelter1980,Straws1985}; however, a majority of theoretical studies have predicted a quintet $^5A_\text{g}$ ground state\cite{Choe1998,Choe1999,Pierloot2009,Hachmann2007,Radon2008,Hajgato2009,Vancoillie2011,Manni2016,Phung2016} instead. It is important to note that here and in previous theoretical studies the calculations were performed on a Fe(P) model cluster. Although the model cluster is not identical to the one studied experimentally, their electronic structures are assumed to be very similar. This assumption needs further examination but for this work we will not pursue this further.

The theoretically predicted quintet ground state may be an artifact of the calculation protocol, such as the size of the active space, the method used to calculate the dynamical correlation or the basis set. Here we will explore the effect of the active space and basis set size on the calculated quintet-triplet gap, while we will leave the exploration of the effect of the dynamical correlation method for future work.

The largest CASSCF-like (FCIQMC-CASSCF) calculations performed to date on Fe(P) model cluster have been those by Li Manni et al.~\cite{Manni2016} which fully correlated 32 electrons in a space of 29 orbitals including the 20 C $2_\text{pz}$, 4 N $2_\text{pz}$, and all 5 Fe $3_d$ orbitals. Results from RASSCF and RASPT2 calculations\cite{Vancoillie2011} have suggested that using a second set of Fe $d$ orbitals can ease the electron repulsion in occupied $d$ orbitals and lower the triplet state energy. This
effect was also observed in a DMRG calculation\cite{Olivares-Amaya2015} where the ground state triplet and quintet calculations were performed using a large active space (44o, 44e) obtained from a $^5A_\text{g}$ Hartree-Fock calculation. This large active space included the 29 orbitals of Li Manni in addition to 15 additional orbitals including the 5 Fe $4_d$, 1 Fe $4_\text{px}$, 1 Fe $4_\text{py}$, 3 N $2_\text{px}$, and 3 N $2_\text{py}$ orbitals. However, picking these orbitals from the results of a canonical Hartree-Fock calculation is non-trivial. Unlike in the original work, here we perform orbital optimization to minimize the effect of the original orbital choice.

We perform HCISCF calculations using two different active spaces, the (29o, 32e) active space of Li Manni and the (44o, 44e) active space used by Olivares-Amaya in the DMRG calculations. Two different HCISCF calculations with (29o, 32e) active space were performed, one with the cc-pVDZ basis set and another with the cc-pVTZ basis set. All calculations were performed using the optimized triplet structure from Ref.~\citenum{Groenhof2005}, which is also the structure used in the DMRG study.

We begin by studying the effect of the accuracy of the HCI calculations in HCISCF on the final optimized active space obtained. Similar to pentacene, three different HCISCF calculations, two of which were vHCISCF with different values of the $\epsilon_1$ threshold and the third was a full HCISCF calculation in which the effect of the PT correction was included. The results, summarized in Table~\ref{FeP_1}, show that the active space orbitals converge much more quickly than the HCISCF energies do. This observation is in agreement with the pentacene results.

\begin{table}[h]
	\caption{\label{FeP_1}
  vHCISCF and HCISCF calculations performed on Fe(P) using various thresholds of $\epsilon_1$. The PT calculation in HCISCF was performed with $\epsilon_2=10^{-5}$ Ha. We also report extrapolated SHCI energies (see text) using the optimized active space orbitals obtained from the SCF calculations. It is interesting to note that (in agreement with pentacene results) although the HCISCF energies are far from converged, the SHCI energies agreement to within 0.1 mHa indicates that the active space orbitals are most likely converged.
	 }

	\renewcommand{\arraystretch}{1.2}
	\begin{tabular}{ P{0.12\columnwidth}P{0.1\columnwidth} P{0.2\columnwidth} P{0.25\columnwidth} P{0.2\columnwidth} }
		\hline\hline
     	&&	$\epsilon_1$ (Ha)	&	$E_\text{HCISCF}$ (Ha)	&	$E_\text{SHCI}$ (Ha)\\
    \hline
    vHCISCF	&&	1$\times 10^{-4}$	&	-2244.9980	&	-2245.0314	 \\
    vHCISCF	&&	5$\times 10^{-5}$	&	-2245.0121	&	-2245.0313 \\
    HCISCF	&&	5$\times 10^{-5}$	&	-2245.0178	&	-2245.0314 \\
		\hline\hline
	\end{tabular}
	\renewcommand{\arraystretch}{1}
\end{table}

Table~\ref{feP_data} summarizes the results of our calculations performed with different active spaces and basis set. For each calculation, we show the vHCISCF energy and the converged HCI energy extrapolated to the FCI limit. The extrapolation procedure is identical to the one used for the butadiene and pentacene calculations, with the largest calculation performed with $\epsilon_1 = 10^{-5}$ Ha. The vHCISCF energies are themselves not important because they are quite far from convergence; however, based on the results shown in Table~\ref{FeP_1} we expect the active space to be converged.

\begin{table*}
	\caption{\label{feP_data} Calculated energies for the Fe(porphyrin) with active spaces of (29o, 32e) and (44o, 44e) and cc-pVDZ and cc-pVTZ basis sets. An initial vHCISCF calculation was performed with $\epsilon_1=10^{-4}$ Ha. The optimized active space was used to perform a more accurate SHCI calculation which was extrapolated to near FCI energies with an estimated error shown in the table. The error bar is calculated as 25\% of the difference between the extrapolated energy and the most
    accurate SHCI energy. $E_\text{ex}$ is the vertical excitation, and $T_\text{OO}$ and $T_\text{CI}$ are the wall time in seconds needed for the orbital optimization and HCI calculation steps during a single vHCISCF iteration, using a single node with two 14-core 2.4 GHz Intel$^\circledR$ Broadwell processors and a combined memory of 128 GB RAM.}

	\renewcommand{\arraystretch}{1.2}
	\begin{tabular}{ c P{0.2\columnwidth} P{0.2\columnwidth} P{0.25\columnwidth} P{0.15\columnwidth}   P{0.2\columnwidth} P{0.15\columnwidth}}
		\hline\hline
		Basis & Sym. & $E_\text{vHCISCF}$ & $E_\text{SHCI}$ & $E_\text{ex}$ & $T_\text{OO}$ & $T_\text{CI}$ \\
		 & & (Ha) & (Ha) & (kcal/mol)  & (sec) & (sec) \\
		\hline
		\multicolumn{7}{c}{}\\
		\multicolumn{7}{c}{CAS(29o, 32e)}\\
        cc-pVDZ & $^5A_\text{g}$    & -2244.9980 & -2245.0314(5) & \multirow{2}{*}{16.7} & 126  & 52  \\
		cc-pVDZ & $^3B_\text{1g}$ & -2244.9776 & -2245.0049(6) &                       & 114 & 56  \\

		\multicolumn{7}{c}{}\\
        cc-pVTZ & $^5A_\text{g}$    & -2245.2229  & -2245.2549(5)& \multirow{2}{*}{16.4} & 2236& 70  \\
		cc-pVTZ & $^3B_\text{1g}$ & -2245.1958 & -2245.2288(6) &                       & 2270& 98  \\

		\multicolumn{7}{c}{}\\
		\multicolumn{7}{c}{CAS(44o,44e)}\\
        cc-pVDZ & $^5A_\text{g}$    & -2245.1457 & -2245.1964(9) & \multirow{2}{*}{-2.0}  & 277  & 185 \\
		cc-pVDZ & $^3B_\text{1g}$ & -2245.1567 & -2245.1995(6) &                       & 264 & 147 \\
		\hline\hline
	\end{tabular}
	\renewcommand{\arraystretch}{1}
\end{table*}

With the smaller (29o, 32e) active space we observe that the quintet state is lower in energy than the triplet state by more than 16 kcal/mol. This result remains virtually unchanged as we go from the cc-pVDZ basis set to the cc-pVTZ basis set. However, when the active space is enlarged to (44o, 44e) we see a switching of the energy ordering and find that the triplet is the ground state. It is worth mentioning that identifying the additional 5 Fe $4_d$, 1 Fe $4_\text{px}$, 1 Fe $4_\text{py}$, 3 N
$2_\text{px}$,
and 3 N $2_\text{py}$ orbitals from the canonical Hartree-Fock orbitals is not trivial and so instead of trying to pick the orbitals by visual inspection we have chosen to include the appropriate number of orbitals (six $A_\text{g}$, three $B_\text{3u}$,  three $B_\text{2u}$, four $B_\text{1g}$, seven $B_\text{1u}$, eight $B_\text{2g}$, eight $B_\text{3g}$ and five $A_\text{u}$ orbitals) from each irreducible representation in the active space.
For both the $^5A_\text{g}$ and the $^3B_\text{1g}$ states, a Hartree-Fock calculation was performed to target the $^5A_\text{1g}$ state, after which the initial guess of the active space was chosen. Unlike in Ref.~\citenum{Olivares-Amaya2015}, at the first iteration of the HCISCF calculations we observe that the $^5A_\text{1g}$ is still the lower energy state. Further, the CASCI energies of the $^5A_\text{1g}$ and the $^3B_\text{1g}$ states were -2244.94423 and -2244.90676 Hartree when
calculated using just the variational HCI with a small $\epsilon_1=10^{-4}$. Although these energies are approximate upper bound of the true energies, they are still lower than the nearly converged DMRG energies reported in Ref.~\citenum{Olivares-Amaya2015}, indicating that our initial active-space orbitals are more appropriate. Although at the first iteration the  $^5A_\text{1g}$  is lower in energy than the  $^3B_\text{1g}$ state, we observe that after HCISCF convergence this ordering is reversed and we obtain the results shown in Table~\ref{feP_data}.

Our results strongly suggest that past theoretical results disagreed with experiments because an insufficiently large active space was used, or an inaccurate method for including dynamical correlation was used. The active space suggested by chemical intuition would not include the high lying virtual orbitals such as the 5 Fe $4_d$, since it should be possible to capture the energy relaxation due to these orbitals with a dynamical correlation method. Such methods are currently being developed in our group and we plan to use these methods with the smaller active space to see if the correct spin ordering can be obtained.

\section{Conclusions}
\label{conclusion}
 The results presented in this work can be used to draw the following three conclusions. First, by using the Lagrangian formulation we can calculate the relaxed reduced density matrices which allow us to straightforwardly use the CASSCF program in PySCF to perform HCISCF. Second, the converged active space orbitals obtained from HCISCF are relatively insensitive to the accuracy of the HCI calculation and consequently loose $\epsilon_1$ and $\epsilon_2$ thresholds can be used. Third, for large active spaces where getting converged HCI energies become difficult, an initial aHCISCF calculation can be performed to optimizes the active space orbitals while keeping the active space itself fixed. These optimized orbitals can vastly improve the convergence of the HCI calculations to the Full CI or CASCI limit, resulting in great speedups.

Here we have exclusively focused on the development of the HCISCF method as a cheap and accurate approximation to CASSCF. To get quantitatively accurate results it is essential to include dynamical correlation effects by allowing excitations outside of the active space. We have recently worked on developing a particularly accurate method for calculating the dynamical correlation called the multireference linearized coupled cluster theory (MRLCC)\cite{Sharma2015,Sharma2016d,Jeanmairet2017}. MRLCC is formulated as a perturbation theory and uses the Fink's partitioning\cite{Fink2006, Fink2009} of the Hamiltonian. We are currently working on combining the internally-contracted MRLCC with the HCISCF calculation, which will be the focus of a forthcoming paper.

\section*{Acknowledgements}
This work was supported through the startup package of the University of Colorado, Boulder. We would also like to thank Cyrus Umrigar for carefully reading the manuscript and suggesting several improvements.

\bibliographystyle{biochem}
\bibliography{hci,paper}

\providecommand{\latin}[1]{#1}
\providecommand*\mcitethebibliography{\thebibliography}
\csname @ifundefined\endcsname{endmcitethebibliography}
  {\let\endmcitethebibliography\endthebibliography}{}
\begin{mcitethebibliography}{103}
\providecommand*\natexlab[1]{#1}
\providecommand*\mciteSetBstSublistMode[1]{}
\providecommand*\mciteSetBstMaxWidthForm[2]{}
\providecommand*\mciteBstWouldAddEndPuncttrue
  {\def\EndOfBibitem{\unskip.}}
\providecommand*\mciteBstWouldAddEndPunctfalse
  {\let\EndOfBibitem\relax}
\providecommand*\mciteSetBstMidEndSepPunct[3]{}
\providecommand*\mciteSetBstSublistLabelBeginEnd[3]{}
\providecommand*\EndOfBibitem{}
\mciteSetBstSublistMode{f}
\mciteSetBstMaxWidthForm{subitem}{(\alph{mcitesubitemcount})}
\mciteSetBstSublistLabelBeginEnd
  {\mcitemaxwidthsubitemform\space}
  {\relax}
  {\relax}

\bibitem[Hohenberg and Kohn(1964)Hohenberg, and
  Kohn]{hohenberg1964inhomogeneous}
Hohenberg,~P., and Kohn,~W. Inhomogeneous electron gas. \emph{Phys. Rev.}
  \textbf{1964} \emph{136}, B864\relax
\mciteBstWouldAddEndPuncttrue
\mciteSetBstMidEndSepPunct{\mcitedefaultmidpunct}
{\mcitedefaultendpunct}{\mcitedefaultseppunct}\relax
\EndOfBibitem
\bibitem[Kohn and Sham(1965)Kohn, and Sham]{kohn1965self}
Kohn,~W., and Sham,~L.~J. Self-consistent equations including exchange and
  correlation effects. \emph{Phys. Rev.} \textbf{1965} \emph{140}, A1133\relax
\mciteBstWouldAddEndPuncttrue
\mciteSetBstMidEndSepPunct{\mcitedefaultmidpunct}
{\mcitedefaultendpunct}{\mcitedefaultseppunct}\relax
\EndOfBibitem
\bibitem[Parr and Weitao(1994)Parr, and Weitao]{parr1994density}
Parr,~R.~G., and Weitao,~Y. \emph{Density-functional theory of atoms and
  molecules}; Oxford university press, 1994; Vol.~16\relax
\mciteBstWouldAddEndPuncttrue
\mciteSetBstMidEndSepPunct{\mcitedefaultmidpunct}
{\mcitedefaultendpunct}{\mcitedefaultseppunct}\relax
\EndOfBibitem
\bibitem[M{\o}ller and Plesset(1934)M{\o}ller, and Plesset]{moller1934note}
M{\o}ller,~C., and Plesset,~M.~S. Note on an approximation treatment for
  many-electron systems. \emph{Physical Review} \textbf{1934} \emph{46},
  618\relax
\mciteBstWouldAddEndPuncttrue
\mciteSetBstMidEndSepPunct{\mcitedefaultmidpunct}
{\mcitedefaultendpunct}{\mcitedefaultseppunct}\relax
\EndOfBibitem
\bibitem[Coester(1958)]{coester1958bound}
Coester,~F. Bound states of a many-particle system. \emph{Nucl. Phys.}
  \textbf{1958} \emph{7}, 421--424\relax
\mciteBstWouldAddEndPuncttrue
\mciteSetBstMidEndSepPunct{\mcitedefaultmidpunct}
{\mcitedefaultendpunct}{\mcitedefaultseppunct}\relax
\EndOfBibitem
\bibitem[{\v{C}}{\'\i}{\v{z}}ek(1966)]{vcivzek1966correlation}
{\v{C}}{\'\i}{\v{z}}ek,~J. On the correlation problem in atomic and molecular
  systems. Calculation of wavefunction components in Ursell-type expansion
  using quantum-field theoretical methods. \emph{J. Chem. Phys.} \textbf{1966}
  \emph{45}, 4256--4266\relax
\mciteBstWouldAddEndPuncttrue
\mciteSetBstMidEndSepPunct{\mcitedefaultmidpunct}
{\mcitedefaultendpunct}{\mcitedefaultseppunct}\relax
\EndOfBibitem
\bibitem[{\v{C}}{\'\i}{\v{z}}ek and Paldus(1980){\v{C}}{\'\i}{\v{z}}ek, and
  Paldus]{cizek1980coupled}
{\v{C}}{\'\i}{\v{z}}ek,~J., and Paldus,~J. Coupled cluster approach.
  \emph{Physica Scripta} \textbf{1980} \emph{21}, 251\relax
\mciteBstWouldAddEndPuncttrue
\mciteSetBstMidEndSepPunct{\mcitedefaultmidpunct}
{\mcitedefaultendpunct}{\mcitedefaultseppunct}\relax
\EndOfBibitem
\bibitem[Purvis~III and Bartlett(1982)Purvis~III, and Bartlett]{purvis1982full}
Purvis~III,~G.~D., and Bartlett,~R.~J. A full coupled-cluster singles and
  doubles model: the inclusion of disconnected triples. \emph{J. Chem. Phys}
  \textbf{1982} \emph{76}, 1910--1918\relax
\mciteBstWouldAddEndPuncttrue
\mciteSetBstMidEndSepPunct{\mcitedefaultmidpunct}
{\mcitedefaultendpunct}{\mcitedefaultseppunct}\relax
\EndOfBibitem
\bibitem[Roos \latin{et~al.}(1980)Roos, Taylor, and
  Siegbahn]{roos1980complete1}
Roos,~B.~O., Taylor,~P.~R., and Siegbahn,~P.~E. A complete active space SCF
  method (CASSCF) using a density matrix formulated super-CI approach.
  \emph{Chem. Phys.} \textbf{1980} \emph{48}, 157--173\relax
\mciteBstWouldAddEndPuncttrue
\mciteSetBstMidEndSepPunct{\mcitedefaultmidpunct}
{\mcitedefaultendpunct}{\mcitedefaultseppunct}\relax
\EndOfBibitem
\bibitem[Roos(1980)]{roos1980complete2}
Roos,~B.~O. The complete active space SCF method in a fock-matrix-based
  super-CI formulation. \emph{International Journal of Quantum Chemistry}
  \textbf{1980} \emph{18}, 175--189\relax
\mciteBstWouldAddEndPuncttrue
\mciteSetBstMidEndSepPunct{\mcitedefaultmidpunct}
{\mcitedefaultendpunct}{\mcitedefaultseppunct}\relax
\EndOfBibitem
\bibitem[Siegbahn \latin{et~al.}(1981)Siegbahn, Alml{\"o}f, Heiberg, and
  Roos]{siegbahn1981complete}
Siegbahn,~P.~E., Alml{\"o}f,~J., Heiberg,~A., and Roos,~B.~O. The complete
  active space SCF (CASSCF) method in a Newton--Raphson formulation with
  application to the HNO molecule. \emph{J. Chem. Phys} \textbf{1981}
  \emph{74}, 2384--2396\relax
\mciteBstWouldAddEndPuncttrue
\mciteSetBstMidEndSepPunct{\mcitedefaultmidpunct}
{\mcitedefaultendpunct}{\mcitedefaultseppunct}\relax
\EndOfBibitem
\bibitem[Roos(2007)]{roos2007complete}
Roos,~B.~O. The Complete Active Space Self-Consistent Field Method and its
  Applications in Electronic Structure Calculations. \emph{Advances in Chemical
  Physics: Ab Initio Methods in Quantum Chemistry Part 2, Volume 69}
  \textbf{2007} 399--445\relax
\mciteBstWouldAddEndPuncttrue
\mciteSetBstMidEndSepPunct{\mcitedefaultmidpunct}
{\mcitedefaultendpunct}{\mcitedefaultseppunct}\relax
\EndOfBibitem
\bibitem[Vogiatzis \latin{et~al.}(2017)Vogiatzis, Ma, Olsen, Gagliardi, and
  Jong]{cas}
Vogiatzis,~K.~D., Ma,~D., Olsen,~J., Gagliardi,~L., and Jong,~W.~d. {Pushing
  Configuration-Interaction to the Limit: Towards Massively Parallel MCSCF
  Calculations}. \emph{arXiv} \textbf{2017} \relax
\mciteBstWouldAddEndPunctfalse
\mciteSetBstMidEndSepPunct{\mcitedefaultmidpunct}
{}{\mcitedefaultseppunct}\relax
\EndOfBibitem
\bibitem[White(1992)]{White1992}
White,~S.~R. {Density matrix formulation for quantum renormalization groups}.
  \emph{Phys. Rev. Lett.} \textbf{1992} \emph{69}, 2863\relax
\mciteBstWouldAddEndPuncttrue
\mciteSetBstMidEndSepPunct{\mcitedefaultmidpunct}
{\mcitedefaultendpunct}{\mcitedefaultseppunct}\relax
\EndOfBibitem
\bibitem[White(1993)]{White1993}
White,~S.~R. {Density-matrix algorithms for quantum renormalization groups}.
  \emph{Phys. Rev. B} \textbf{1993} \emph{48}, 10345\relax
\mciteBstWouldAddEndPuncttrue
\mciteSetBstMidEndSepPunct{\mcitedefaultmidpunct}
{\mcitedefaultendpunct}{\mcitedefaultseppunct}\relax
\EndOfBibitem
\bibitem[Fano \latin{et~al.}(1998)Fano, Ortolani, and Ziosi]{Fano1998}
Fano,~G., Ortolani,~F., and Ziosi,~L. {The density matrix renormalization group
  method: Application to the PPP model of a cyclic polyene chain}.
  \emph{Journal of Chemical Physics} \textbf{1998} \emph{108}, 9246--9252\relax
\mciteBstWouldAddEndPuncttrue
\mciteSetBstMidEndSepPunct{\mcitedefaultmidpunct}
{\mcitedefaultendpunct}{\mcitedefaultseppunct}\relax
\EndOfBibitem
\bibitem[White and Martin(1999)White, and Martin]{White1999}
White,~S.~R., and Martin,~R.~L. {Ab initio quantum chemistry using the density
  matrix renormalization group}. \emph{J. Chem. Phys.} \textbf{1999}
  \emph{110}, 4127\relax
\mciteBstWouldAddEndPuncttrue
\mciteSetBstMidEndSepPunct{\mcitedefaultmidpunct}
{\mcitedefaultendpunct}{\mcitedefaultseppunct}\relax
\EndOfBibitem
\bibitem[Schollw\"ock(2005)]{sch05}
Schollw\"ock,~U. The density-matrix renormalization group. \emph{Rev. Mod.
  Phys.} \textbf{2005} \emph{77}, 259--315\relax
\mciteBstWouldAddEndPuncttrue
\mciteSetBstMidEndSepPunct{\mcitedefaultmidpunct}
{\mcitedefaultendpunct}{\mcitedefaultseppunct}\relax
\EndOfBibitem
\bibitem[Szalay \latin{et~al.}(2015)Szalay, Pfeffer, Murg, Barcza, Verstraete,
  Schneider, and Legeza]{legeza2015}
Szalay,~S., Pfeffer,~M., Murg,~V., Barcza,~G., Verstraete,~F., Schneider,~R.,
  and Legeza,~Ã. Tensor product methods and entanglement optimization for ab
  initio quantum chemistry. \emph{International Journal of Quantum Chemistry}
  \textbf{2015} \emph{115}, 1342--1391\relax
\mciteBstWouldAddEndPuncttrue
\mciteSetBstMidEndSepPunct{\mcitedefaultmidpunct}
{\mcitedefaultendpunct}{\mcitedefaultseppunct}\relax
\EndOfBibitem
\bibitem[Schollwöck(2011)]{sch11}
Schollwöck,~U. The density-matrix renormalization group in the age of matrix
  product states. \emph{Annals of Physics} \textbf{2011} \emph{326}, 96 -- 192,
  January 2011 Special Issue\relax
\mciteBstWouldAddEndPuncttrue
\mciteSetBstMidEndSepPunct{\mcitedefaultmidpunct}
{\mcitedefaultendpunct}{\mcitedefaultseppunct}\relax
\EndOfBibitem
\bibitem[Daul \latin{et~al.}(2001)Daul, Ciofini, Daul, and White]{Daul2001}
Daul,~S., Ciofini,~I., Daul,~C., and White,~S.~R. {Quantum chemistry using the
  density matrix renormalization group}. \emph{J. Chem. Phys.} \textbf{2001}
  \emph{115}, 6815--6821\relax
\mciteBstWouldAddEndPuncttrue
\mciteSetBstMidEndSepPunct{\mcitedefaultmidpunct}
{\mcitedefaultendpunct}{\mcitedefaultseppunct}\relax
\EndOfBibitem
\bibitem[Chan and Head-Gordon(2002)Chan, and Head-Gordon]{Chan2002}
Chan,~G. K.~L., and Head-Gordon,~M. {Highly correlated calculations with a
  polynomial cost algorithm: A study of the density matrix renormalization
  group}. \emph{J. Chem. Phys.} \textbf{2002} \emph{116}, 4462\relax
\mciteBstWouldAddEndPuncttrue
\mciteSetBstMidEndSepPunct{\mcitedefaultmidpunct}
{\mcitedefaultendpunct}{\mcitedefaultseppunct}\relax
\EndOfBibitem
\bibitem[Moritz and Reiher(2006)Moritz, and Reiher]{Moritz2006}
Moritz,~G., and Reiher,~M. {Construction of environment states in
  quantum-chemical density-matrix renormalization group calculations}.
  \emph{Journal of Chemical Physics} \textbf{2006} \emph{124}, 1--9\relax
\mciteBstWouldAddEndPuncttrue
\mciteSetBstMidEndSepPunct{\mcitedefaultmidpunct}
{\mcitedefaultendpunct}{\mcitedefaultseppunct}\relax
\EndOfBibitem
\bibitem[Zgid and Nooijen(2008)Zgid, and Nooijen]{Zgid2008}
Zgid,~D., and Nooijen,~M. {On the spin and symmetry adaptation of the density
  matrix renormalization group method.} \emph{J. Chem. Phys.} \textbf{2008}
  \emph{128}, 014107\relax
\mciteBstWouldAddEndPuncttrue
\mciteSetBstMidEndSepPunct{\mcitedefaultmidpunct}
{\mcitedefaultendpunct}{\mcitedefaultseppunct}\relax
\EndOfBibitem
\bibitem[Luo \latin{et~al.}(2010)Luo, Qin, and Xiang]{Luo2010}
Luo,~H.~G., Qin,~M.~P., and Xiang,~T. {Optimizing Hartree-Fock orbitals by the
  density-matrix renormalization group}. \emph{Physical Review B - Condensed
  Matter and Materials Physics} \textbf{2010} \emph{81}, 1--4\relax
\mciteBstWouldAddEndPuncttrue
\mciteSetBstMidEndSepPunct{\mcitedefaultmidpunct}
{\mcitedefaultendpunct}{\mcitedefaultseppunct}\relax
\EndOfBibitem
\bibitem[Marti and Reiher(2010)Marti, and Reiher]{Marti2010}
Marti,~K.~H., and Reiher,~M. {The Density Matrix Renormalization Group
  Algorithm in Quantum Chemistry}. \emph{Zeitschrift f{\"{u}}r Physikalische
  Chemie} \textbf{2010} \emph{224}, 583--599\relax
\mciteBstWouldAddEndPuncttrue
\mciteSetBstMidEndSepPunct{\mcitedefaultmidpunct}
{\mcitedefaultendpunct}{\mcitedefaultseppunct}\relax
\EndOfBibitem
\bibitem[Chan and Sharma(2011)Chan, and Sharma]{Chan2011}
Chan,~G. K.-L., and Sharma,~S. {The Density Matrix Renormalization Group in
  Quantum Chemistry}. \emph{Ann. Rev. Phys. Chem.} \textbf{2011} \emph{62},
  465\relax
\mciteBstWouldAddEndPuncttrue
\mciteSetBstMidEndSepPunct{\mcitedefaultmidpunct}
{\mcitedefaultendpunct}{\mcitedefaultseppunct}\relax
\EndOfBibitem
\bibitem[Kurashige and Yanai(2011)Kurashige, and Yanai]{Kurashige2011}
Kurashige,~Y., and Yanai,~T. {Second-order perturbation theory with a density
  matrix renormalization group self-consistent field reference function: theory
  and application to the study of chromium dimer.} \emph{J. Chem. Phys.}
  \textbf{2011} \emph{135}, 094104\relax
\mciteBstWouldAddEndPuncttrue
\mciteSetBstMidEndSepPunct{\mcitedefaultmidpunct}
{\mcitedefaultendpunct}{\mcitedefaultseppunct}\relax
\EndOfBibitem
\bibitem[Sharma and Chan(2012)Sharma, and Chan]{sharmaspin}
Sharma,~S., and Chan,~G. K.-L. {Spin-adapted density matrix renormalization
  group algorithms for quantum chemistry}. \emph{J. Chem. Phys.} \textbf{2012}
  \emph{136}, 124121\relax
\mciteBstWouldAddEndPuncttrue
\mciteSetBstMidEndSepPunct{\mcitedefaultmidpunct}
{\mcitedefaultendpunct}{\mcitedefaultseppunct}\relax
\EndOfBibitem
\bibitem[Sharma \latin{et~al.}(2014)Sharma, Sivalingam, Neese, and
  Chan]{sha-nat}
Sharma,~S., Sivalingam,~K., Neese,~F., and Chan,~G. K.-L. Low-energy spectrum
  of iron–sulfur clusters directly from many-particle quantum mechanics.
  \emph{Nature Chemistry} \textbf{2014} 927\relax
\mciteBstWouldAddEndPuncttrue
\mciteSetBstMidEndSepPunct{\mcitedefaultmidpunct}
{\mcitedefaultendpunct}{\mcitedefaultseppunct}\relax
\EndOfBibitem
\bibitem[Kurashige \latin{et~al.}(2013)Kurashige, Chan, and Yanai]{kura-nat}
Kurashige,~Y., Chan,~G. K.-L., and Yanai,~T. Entangled quantum electronic
  wavefunctions of the Mn4CaO5 cluster in photosystem II. \emph{Nature
  Chemistry} \textbf{2013} 660\relax
\mciteBstWouldAddEndPuncttrue
\mciteSetBstMidEndSepPunct{\mcitedefaultmidpunct}
{\mcitedefaultendpunct}{\mcitedefaultseppunct}\relax
\EndOfBibitem
\bibitem[Wouters \latin{et~al.}(2014)Wouters, Bogaerts, {Van Der Voort}, {Van
  Speybroeck}, and {Van Neck}]{wouters2014}
Wouters,~S., Bogaerts,~T., {Van Der Voort},~P., {Van Speybroeck},~V., and {Van
  Neck},~D. {Communication: DMRG-SCF study of the singlet, triplet, and quintet
  states of oxo-Mn(Salen).} \emph{J. Chem. Phys.} \textbf{2014} \emph{140},
  241103\relax
\mciteBstWouldAddEndPuncttrue
\mciteSetBstMidEndSepPunct{\mcitedefaultmidpunct}
{\mcitedefaultendpunct}{\mcitedefaultseppunct}\relax
\EndOfBibitem
\bibitem[Keller and Reiher(2016)Keller, and Reiher]{keller}
Keller,~S., and Reiher,~M. Spin-adapted matrix product states and operators.
  \emph{The Journal of Chemical Physics} \textbf{2016} \emph{144}, 134101\relax
\mciteBstWouldAddEndPuncttrue
\mciteSetBstMidEndSepPunct{\mcitedefaultmidpunct}
{\mcitedefaultendpunct}{\mcitedefaultseppunct}\relax
\EndOfBibitem
\bibitem[Kurashige(2014)]{yuki-review}
Kurashige,~Y. Multireference electron correlation methods with density matrix
  renormalisation group reference functions. \emph{Molecular Physics}
  \textbf{2014} \emph{112}, 1485--1494\relax
\mciteBstWouldAddEndPuncttrue
\mciteSetBstMidEndSepPunct{\mcitedefaultmidpunct}
{\mcitedefaultendpunct}{\mcitedefaultseppunct}\relax
\EndOfBibitem
\bibitem[Yanai \latin{et~al.}(2015)Yanai, Kurashige, Mizukami, Chalupský, Lan,
  and Saitow]{takeshi15}
Yanai,~T., Kurashige,~Y., Mizukami,~W., Chalupský,~J., Lan,~T.~N., and
  Saitow,~M. Density matrix renormalization group for ab initio Calculations
  and associated dynamic correlation methods: A review of theory and
  applications. \emph{International Journal of Quantum Chemistry} \textbf{2015}
  \emph{115}, 283--299\relax
\mciteBstWouldAddEndPuncttrue
\mciteSetBstMidEndSepPunct{\mcitedefaultmidpunct}
{\mcitedefaultendpunct}{\mcitedefaultseppunct}\relax
\EndOfBibitem
\bibitem[Malmqvist \latin{et~al.}(1990)Malmqvist, Rendell, and
  Roos]{malmqvist1990restricted}
Malmqvist,~P.~A., Rendell,~A., and Roos,~B.~O. The restricted active space
  self-consistent-field method, implemented with a split graph unitary group
  approach. \emph{J. Phys. Chem.} \textbf{1990} \emph{94}, 5477--5482\relax
\mciteBstWouldAddEndPuncttrue
\mciteSetBstMidEndSepPunct{\mcitedefaultmidpunct}
{\mcitedefaultendpunct}{\mcitedefaultseppunct}\relax
\EndOfBibitem
\bibitem[Celani and Werner(2000)Celani, and Werner]{celani2000multireference}
Celani,~P., and Werner,~H.-J. Multireference perturbation theory for large
  restricted and selected active space reference wave functions. \emph{J. Chem.
  Phys.} \textbf{2000} \emph{112}, 5546--5557\relax
\mciteBstWouldAddEndPuncttrue
\mciteSetBstMidEndSepPunct{\mcitedefaultmidpunct}
{\mcitedefaultendpunct}{\mcitedefaultseppunct}\relax
\EndOfBibitem
\bibitem[Ma \latin{et~al.}(2011)Ma, Li~Manni, and Gagliardi]{ma2011generalized}
Ma,~D., Li~Manni,~G., and Gagliardi,~L. The generalized active space concept in
  multiconfigurational self-consistent field methods. \emph{J. Chem. Phys.}
  \textbf{2011} \emph{135}, 044128\relax
\mciteBstWouldAddEndPuncttrue
\mciteSetBstMidEndSepPunct{\mcitedefaultmidpunct}
{\mcitedefaultendpunct}{\mcitedefaultseppunct}\relax
\EndOfBibitem
\bibitem[Nakata \latin{et~al.}(2002)Nakata, Ehara, and
  Nakatsuji]{NakEhaNak-JCP-02}
Nakata,~M., Ehara,~M., and Nakatsuji,~H. Density matrix variational theory:
  Application to the potential energy surfaces and strongly correlated systems.
  \emph{J. Chem. Phys.} \textbf{2002} \emph{116}, 5432\relax
\mciteBstWouldAddEndPuncttrue
\mciteSetBstMidEndSepPunct{\mcitedefaultmidpunct}
{\mcitedefaultendpunct}{\mcitedefaultseppunct}\relax
\EndOfBibitem
\bibitem[Mazziotti(2006)]{mazziotti2006quantum}
Mazziotti,~D.~A. Quantum chemistry without wave functions: two-electron reduced
  density matrices. \emph{Accounts of chemical research} \textbf{2006}
  \emph{39}, 207--215\relax
\mciteBstWouldAddEndPuncttrue
\mciteSetBstMidEndSepPunct{\mcitedefaultmidpunct}
{\mcitedefaultendpunct}{\mcitedefaultseppunct}\relax
\EndOfBibitem
\bibitem[Valdemoro({2007})]{Val-ACP-07}
Valdemoro,~C. In \emph{{Reduced-density-matrix mechanics with applications to
  many-electron atoms and molecules}}; Mazziotti,~D.~A., Ed.; {Adv. Chem.
  Phys.}; {2007}; Vol. {134}; pp {121--164}\relax
\mciteBstWouldAddEndPuncttrue
\mciteSetBstMidEndSepPunct{\mcitedefaultmidpunct}
{\mcitedefaultendpunct}{\mcitedefaultseppunct}\relax
\EndOfBibitem
\bibitem[Ivanic and Ruedenberg(2001)Ivanic, and Ruedenberg]{Ivanic2001}
Ivanic,~J., and Ruedenberg,~K. {Identification of deadwood in configuration
  spaces through general direct configuration interaction}. \emph{Theoretical
  Chemistry Accounts} \textbf{2001} \emph{106}, 339--351\relax
\mciteBstWouldAddEndPuncttrue
\mciteSetBstMidEndSepPunct{\mcitedefaultmidpunct}
{\mcitedefaultendpunct}{\mcitedefaultseppunct}\relax
\EndOfBibitem
\bibitem[Huron(1973)]{Huron1973}
Huron,~B. {Iterative perturbation calculations of ground and excited state
  energies from multiconfigurational zeroth-order wavefunctions}. \emph{J.
  Chem. Phys.} \textbf{1973} \emph{58}, 5745\relax
\mciteBstWouldAddEndPuncttrue
\mciteSetBstMidEndSepPunct{\mcitedefaultmidpunct}
{\mcitedefaultendpunct}{\mcitedefaultseppunct}\relax
\EndOfBibitem
\bibitem[Buenker and Peyerimhoff(1974)Buenker, and Peyerimhoff]{Buenker1974}
Buenker,~R.~J., and Peyerimhoff,~S.~D. {Individualized configuration selection
  in CI calculations with subsequent energy extrapolation}. \emph{Theoretica
  chimica acta} \textbf{1974} \emph{35}, 33--58\relax
\mciteBstWouldAddEndPuncttrue
\mciteSetBstMidEndSepPunct{\mcitedefaultmidpunct}
{\mcitedefaultendpunct}{\mcitedefaultseppunct}\relax
\EndOfBibitem
\bibitem[Evangelista \latin{et~al.}(1983)Evangelista, Daudey, and
  Malrieu]{Evangelista1983}
Evangelista,~F.~A., Daudey,~J.~P., and Malrieu,~J.~P. {CONVERGENCE OF AN
  IMPROVED ClPSl ALGORITHM}. \emph{Chemical Physics} \textbf{1983} \emph{75},
  91--102\relax
\mciteBstWouldAddEndPuncttrue
\mciteSetBstMidEndSepPunct{\mcitedefaultmidpunct}
{\mcitedefaultendpunct}{\mcitedefaultseppunct}\relax
\EndOfBibitem
\bibitem[Harrison(1991)]{Harrison1991}
Harrison,~R.~J. {Approximating full configuration interaction with selected
  configuration interaction and perturbation theory}. \emph{The Journal of
  Chemical Physics} \textbf{1991} \emph{94}, 5021--5031\relax
\mciteBstWouldAddEndPuncttrue
\mciteSetBstMidEndSepPunct{\mcitedefaultmidpunct}
{\mcitedefaultendpunct}{\mcitedefaultseppunct}\relax
\EndOfBibitem
\bibitem[Steiner \latin{et~al.}(1994)Steiner, Wenzel, Wilson, and
  Wilkins]{Steiner1994}
Steiner,~M.~M., Wenzel,~W., Wilson,~K.~G., and Wilkins,~J.~W. {The efficient
  treatment of higher excitations in CI calculations. A comparison of exact and
  approximate results}. \emph{Chemical Physics Letters} \textbf{1994}
  \emph{231}, 263--268\relax
\mciteBstWouldAddEndPuncttrue
\mciteSetBstMidEndSepPunct{\mcitedefaultmidpunct}
{\mcitedefaultendpunct}{\mcitedefaultseppunct}\relax
\EndOfBibitem
\bibitem[Wenzel \latin{et~al.}(1996)Wenzel, Steiner, and Wilson]{Wenzel1996}
Wenzel,~W., Steiner,~M.~M., and Wilson,~K.~G. {Multireference Basis-Set
  Reduction}. \emph{International Journal of Quantum Chemistry} \textbf{1996}
  \emph{30}, 1325--1330\relax
\mciteBstWouldAddEndPuncttrue
\mciteSetBstMidEndSepPunct{\mcitedefaultmidpunct}
{\mcitedefaultendpunct}{\mcitedefaultseppunct}\relax
\EndOfBibitem
\bibitem[Neese(2003)]{Neese2003}
Neese,~F. {A spectroscopy oriented configuration interaction procedure}.
  \emph{Journal of Chemical Physics} \textbf{2003} \emph{119}, 9428--9443\relax
\mciteBstWouldAddEndPuncttrue
\mciteSetBstMidEndSepPunct{\mcitedefaultmidpunct}
{\mcitedefaultendpunct}{\mcitedefaultseppunct}\relax
\EndOfBibitem
\bibitem[Abrams and Sherrill(2005)Abrams, and Sherrill]{Abrams2005}
Abrams,~M.~L., and Sherrill,~C.~D. {Important configurations in configuration
  interaction and coupled-cluster wave functions}. \emph{Chemical Physics
  Letters} \textbf{2005} \emph{412}, 121--124\relax
\mciteBstWouldAddEndPuncttrue
\mciteSetBstMidEndSepPunct{\mcitedefaultmidpunct}
{\mcitedefaultendpunct}{\mcitedefaultseppunct}\relax
\EndOfBibitem
\bibitem[Bytautas and Ruedenberg(2009)Bytautas, and Ruedenberg]{Bytautas2009}
Bytautas,~L., and Ruedenberg,~K. {A priori identification of configurational
  deadwood}. \emph{Chemical Physics} \textbf{2009} \emph{356}, 64--75\relax
\mciteBstWouldAddEndPuncttrue
\mciteSetBstMidEndSepPunct{\mcitedefaultmidpunct}
{\mcitedefaultendpunct}{\mcitedefaultseppunct}\relax
\EndOfBibitem
\bibitem[Evangelista(2014)]{Evangelista2014}
Evangelista,~F.~A. {A driven similarity renormalization group approach to
  quantum many-body problems}. \emph{Journal of Chemical Physics} \textbf{2014}
  \emph{141}\relax
\mciteBstWouldAddEndPuncttrue
\mciteSetBstMidEndSepPunct{\mcitedefaultmidpunct}
{\mcitedefaultendpunct}{\mcitedefaultseppunct}\relax
\EndOfBibitem
\bibitem[Knowles(2015)]{Knowles2015}
Knowles,~P.~J. {Compressive sampling in configuration interaction
  wavefunctions}. \emph{Molecular Physics} \textbf{2015} \emph{113},
  1655--1660\relax
\mciteBstWouldAddEndPuncttrue
\mciteSetBstMidEndSepPunct{\mcitedefaultmidpunct}
{\mcitedefaultendpunct}{\mcitedefaultseppunct}\relax
\EndOfBibitem
\bibitem[Schriber and Evangelista(2016)Schriber, and Evangelista]{Schriber2016}
Schriber,~J.~B., and Evangelista,~F.~A. {Communication: An adaptive
  configuration interaction approach for strongly correlated electrons with
  tunable accuracy.} \emph{J. Chem. Phys.} \textbf{2016} \emph{144},
  161106\relax
\mciteBstWouldAddEndPuncttrue
\mciteSetBstMidEndSepPunct{\mcitedefaultmidpunct}
{\mcitedefaultendpunct}{\mcitedefaultseppunct}\relax
\EndOfBibitem
\bibitem[Liu and Hoffmann(2016)Liu, and Hoffmann]{Liu2016}
Liu,~W., and Hoffmann,~M.~R. {ICI: Iterative CI toward full CI}. \emph{Journal
  of Chemical Theory and Computation} \textbf{2016} \emph{12}, 1169--1178\relax
\mciteBstWouldAddEndPuncttrue
\mciteSetBstMidEndSepPunct{\mcitedefaultmidpunct}
{\mcitedefaultendpunct}{\mcitedefaultseppunct}\relax
\EndOfBibitem
\bibitem[Caffarel \latin{et~al.}(2016)Caffarel, Applencourt, Giner, and
  Scemama]{Caffarel2016}
Caffarel,~M., Applencourt,~T., Giner,~E., and Scemama,~A. {Using CIPSI Nodes in
  Diffusion Monte Carlo}. \emph{ACS Symposium Series} \textbf{2016}
  \emph{1234}, 15--46\relax
\mciteBstWouldAddEndPuncttrue
\mciteSetBstMidEndSepPunct{\mcitedefaultmidpunct}
{\mcitedefaultendpunct}{\mcitedefaultseppunct}\relax
\EndOfBibitem
\bibitem[Garniron \latin{et~al.}(2017)Garniron, Scemama, Loos, and
  Caffarel]{yann2017}
Garniron,~Y., Scemama,~A., Loos,~P.-F., and Caffarel,~M. Hybrid
  stochastic-deterministic calculation of the second-order perturbative
  contribution of multireference perturbation theory. \emph{The Journal of
  Chemical Physics} \textbf{2017} \emph{147}, 034101\relax
\mciteBstWouldAddEndPuncttrue
\mciteSetBstMidEndSepPunct{\mcitedefaultmidpunct}
{\mcitedefaultendpunct}{\mcitedefaultseppunct}\relax
\EndOfBibitem
\bibitem[Booth \latin{et~al.}(2009)Booth, Thom, and Alavi]{Booth2009}
Booth,~G.~H., Thom,~A. J.~W., and Alavi,~A. {Fermion Monte Carlo without fixed
  nodes: a game of life, death, and annihilation in Slater determinant space.}
  \emph{J. Chem. Phys.} \textbf{2009} \emph{131}, 054106\relax
\mciteBstWouldAddEndPuncttrue
\mciteSetBstMidEndSepPunct{\mcitedefaultmidpunct}
{\mcitedefaultendpunct}{\mcitedefaultseppunct}\relax
\EndOfBibitem
\bibitem[Cleland \latin{et~al.}(2010)Cleland, Booth, and Alavi]{Cleland2010}
Cleland,~D., Booth,~G.~H., and Alavi,~A. {Communications: Survival of the
  fittest: accelerating convergence in full configuration-interaction quantum
  Monte Carlo.} \emph{J. Chem. Phys.} \textbf{2010} \emph{132}, 041103\relax
\mciteBstWouldAddEndPuncttrue
\mciteSetBstMidEndSepPunct{\mcitedefaultmidpunct}
{\mcitedefaultendpunct}{\mcitedefaultseppunct}\relax
\EndOfBibitem
\bibitem[Petruzielo \latin{et~al.}(2012)Petruzielo, Holmes, Changlani,
  Nightingale, and Umrigar]{Petruzielo2012}
Petruzielo,~F.~R., Holmes,~A.~A., Changlani,~H.~J., Nightingale,~M.~P., and
  Umrigar,~C.~J. {Semistochastic projector monte carlo method}. \emph{Physical
  Review Letters} \textbf{2012} \emph{109}, 1--5\relax
\mciteBstWouldAddEndPuncttrue
\mciteSetBstMidEndSepPunct{\mcitedefaultmidpunct}
{\mcitedefaultendpunct}{\mcitedefaultseppunct}\relax
\EndOfBibitem
\bibitem[Thomas \latin{et~al.}(2015)Thomas, Sun, Alavi, and Booth]{thomas2015}
Thomas,~R.~E., Sun,~Q., Alavi,~A., and Booth,~G.~H. Stochastic
  Multiconfigurational Self-Consistent Field Theory. \emph{Journal of Chemical
  Theory and Computation} \textbf{2015} \emph{11}, 5316--5325\relax
\mciteBstWouldAddEndPuncttrue
\mciteSetBstMidEndSepPunct{\mcitedefaultmidpunct}
{\mcitedefaultendpunct}{\mcitedefaultseppunct}\relax
\EndOfBibitem
\bibitem[Holmes \latin{et~al.}({2016})Holmes, Tubman, and
  Umrigar]{HolTubUmr-JCTC-16}
Holmes,~A.~A., Tubman,~N.~M., and Umrigar,~C.~J. Heat-bath Configuration
  Interaction: An efficient selected CI algorithm inspired by heat-bath
  sampling. \emph{J. Chem. Theory Comput.} \textbf{{2016}} \emph{{12}},
  {3674}\relax
\mciteBstWouldAddEndPuncttrue
\mciteSetBstMidEndSepPunct{\mcitedefaultmidpunct}
{\mcitedefaultendpunct}{\mcitedefaultseppunct}\relax
\EndOfBibitem
\bibitem[Sharma \latin{et~al.}(2017)Sharma, Holmes, Jeanmairet, Alavi, and
  Umrigar]{ShaHolUmr-JCTC-17}
Sharma,~S., Holmes,~A.~A., Jeanmairet,~G., Alavi,~A., and Umrigar,~C.~J.
  Semistochastic Heat-bath Configuration Interaction method: selected
  configuration interaction with semistochastic perturbation theory. \emph{J.
  Chem. Theory Comput.} \textbf{2017} \emph{13}, 1595--1604\relax
\mciteBstWouldAddEndPuncttrue
\mciteSetBstMidEndSepPunct{\mcitedefaultmidpunct}
{\mcitedefaultendpunct}{\mcitedefaultseppunct}\relax
\EndOfBibitem
\bibitem[Ma \latin{et~al.}(2017)Ma, Knecht, Keller, and Reiher]{macasscf}
Ma,~Y., Knecht,~S., Keller,~S., and Reiher,~M. Second-Order
  Self-Consistent-Field Density-Matrix Renormalization Group. \emph{Journal of
  Chemical Theory and Computation} \textbf{2017} \emph{13}, 2533--2549\relax
\mciteBstWouldAddEndPuncttrue
\mciteSetBstMidEndSepPunct{\mcitedefaultmidpunct}
{\mcitedefaultendpunct}{\mcitedefaultseppunct}\relax
\EndOfBibitem
\bibitem[Ghosh \latin{et~al.}(2008)Ghosh, Hachmann, Yanai, and Chan]{Ghosh2008}
Ghosh,~D., Hachmann,~J., Yanai,~T., and Chan,~G. K.~L. {Orbital optimization in
  the density matrix renormalization group, with applications to polyenes and
  $\beta$-carotene}. \emph{J. Chem. Phys.} \textbf{2008} \emph{128},
  144117\relax
\mciteBstWouldAddEndPuncttrue
\mciteSetBstMidEndSepPunct{\mcitedefaultmidpunct}
{\mcitedefaultendpunct}{\mcitedefaultseppunct}\relax
\EndOfBibitem
\bibitem[Zgid and Nooijen(2008)Zgid, and Nooijen]{zgidcasscf}
Zgid,~D., and Nooijen,~M. The density matrix renormalization group
  self-consistent field method: Orbital optimization with the density matrix
  renormalization group method in the active space. \emph{The Journal of
  Chemical Physics} \textbf{2008} \emph{128}, 144116\relax
\mciteBstWouldAddEndPuncttrue
\mciteSetBstMidEndSepPunct{\mcitedefaultmidpunct}
{\mcitedefaultendpunct}{\mcitedefaultseppunct}\relax
\EndOfBibitem
\bibitem[Manni \latin{et~al.}(2016)Manni, Smart, and Alavi]{Manni2016}
Manni,~G.~L., Smart,~S.~D., and Alavi,~A. {Combining the Complete Active Space
  Self-Consistent Field Method and the Full Con fi guration Interaction Quantum
  Monte Carlo within a Super-CI Framework , with Application to Challenging
  Metal- Porphyrins}. \emph{Journal of Chemical Theory and Computation}
  \textbf{2016} \emph{12}, 1245--1258\relax
\mciteBstWouldAddEndPuncttrue
\mciteSetBstMidEndSepPunct{\mcitedefaultmidpunct}
{\mcitedefaultendpunct}{\mcitedefaultseppunct}\relax
\EndOfBibitem
\bibitem[Thomas \latin{et~al.}(2015)Thomas, Sun, Alavi, and Booth]{thomas15}
Thomas,~R.~E., Sun,~Q., Alavi,~A., and Booth,~G.~H. Stochastic
  Multiconfigurational Self-Consistent Field Theory. \emph{Journal of Chemical
  Theory and Computation} \textbf{2015} \emph{11}, 5316--5325\relax
\mciteBstWouldAddEndPuncttrue
\mciteSetBstMidEndSepPunct{\mcitedefaultmidpunct}
{\mcitedefaultendpunct}{\mcitedefaultseppunct}\relax
\EndOfBibitem
\bibitem[Lochan and Head-Gordon(2007)Lochan, and Head-Gordon]{Lochan2007}
Lochan,~R.~C., and Head-Gordon,~M. {Orbital-optimized opposite-spin scaled
  second-order correlation: An economical method to improve the description of
  open-shell molecules}. \emph{Journal of Chemical Physics} \textbf{2007}
  \emph{126}\relax
\mciteBstWouldAddEndPuncttrue
\mciteSetBstMidEndSepPunct{\mcitedefaultmidpunct}
{\mcitedefaultendpunct}{\mcitedefaultseppunct}\relax
\EndOfBibitem
\bibitem[Neese \latin{et~al.}(2009)Neese, Schwabe, Kossmann, Schirmer, and
  Grimme]{Neese2009}
Neese,~F., Schwabe,~T., Kossmann,~S., Schirmer,~B., and Grimme,~S. {Assessment
  of orbital-optimized, spin-component scaled second-order many-body
  perturbation theory for thermochemistry and kinetics}. \emph{Journal of
  Chemical Theory and Computation} \textbf{2009} \emph{5}, 3060--3073\relax
\mciteBstWouldAddEndPuncttrue
\mciteSetBstMidEndSepPunct{\mcitedefaultmidpunct}
{\mcitedefaultendpunct}{\mcitedefaultseppunct}\relax
\EndOfBibitem
\bibitem[Burgos \latin{et~al.}(1977)Burgos, Pope, Swenberg, and
  Alfano]{BurgosJ.PopeM.SwenbergC.E.Alfano1977}
Burgos,~J., Pope,~M., Swenberg,~C.~E., and Alfano,~R.~R. {Heterofission in
  pentacene-doped tetracene single crystals}. \emph{Phsica status solidi. B,
  Basis reserach} \textbf{1977} \emph{83}, 249--256\relax
\mciteBstWouldAddEndPuncttrue
\mciteSetBstMidEndSepPunct{\mcitedefaultmidpunct}
{\mcitedefaultendpunct}{\mcitedefaultseppunct}\relax
\EndOfBibitem
\bibitem[Biermann and Schmidt(1980)Biermann, and
  Schmidt]{BiermannD.Schmidt1979}
Biermann,~D., and Schmidt,~W. {Diels-Alder Reactivity of Polycyclic Aromatic
  Hydrocarbons III New Experimental and Theoretical Results}. \emph{Israel
  Journal of Chemistry} \textbf{1980} \emph{20}, 312--318\relax
\mciteBstWouldAddEndPuncttrue
\mciteSetBstMidEndSepPunct{\mcitedefaultmidpunct}
{\mcitedefaultendpunct}{\mcitedefaultseppunct}\relax
\EndOfBibitem
\bibitem[Biermann and Schmidt(1980)Biermann, and Schmidt]{Biermann1980}
Biermann,~D., and Schmidt,~W. {Diels-Alder Reactivity of Polycyclic Aromatic
  Hydrocarbons. 1. Acenes and Benzologs}. \emph{Journal of the American
  Chemical Society} \textbf{1980} \emph{102}, 3163--3173\relax
\mciteBstWouldAddEndPuncttrue
\mciteSetBstMidEndSepPunct{\mcitedefaultmidpunct}
{\mcitedefaultendpunct}{\mcitedefaultseppunct}\relax
\EndOfBibitem
\bibitem[Hachmann \latin{et~al.}(2007)Hachmann, Dorando, Avilés, and
  Chan]{Hachmann2007}
Hachmann,~J., Dorando,~J.~J., Avilés,~M., and Chan,~G. K.-L. The radical
  character of the acenes: A density matrix renormalization group study.
  \emph{The Journal of Chemical Physics} \textbf{2007} \emph{127}, 134309\relax
\mciteBstWouldAddEndPuncttrue
\mciteSetBstMidEndSepPunct{\mcitedefaultmidpunct}
{\mcitedefaultendpunct}{\mcitedefaultseppunct}\relax
\EndOfBibitem
\bibitem[Dorando \latin{et~al.}(2007)Dorando, Hachmann, and Chan]{Dorando2007}
Dorando,~J.~J., Hachmann,~J., and Chan,~G. K.-L. {Targeted excited state
  algorithms}. \emph{J. Chem. Phys.} \textbf{2007} \emph{127}, 84109\relax
\mciteBstWouldAddEndPuncttrue
\mciteSetBstMidEndSepPunct{\mcitedefaultmidpunct}
{\mcitedefaultendpunct}{\mcitedefaultseppunct}\relax
\EndOfBibitem
\bibitem[Hajgat{\'{o}} \latin{et~al.}(2009)Hajgat{\'{o}}, Szieberth, Geerlings,
  {De Proft}, and Deleuze]{Hajgato2009}
Hajgat{\'{o}},~B., Szieberth,~D., Geerlings,~P., {De Proft},~F., and
  Deleuze,~M.~S. {A benchmark theoretical study of the electronic ground state
  and of the singlet-triplet split of benzene and linear acenes}. \emph{Journal
  of Chemical Physics} \textbf{2009} \emph{131}, 1--18\relax
\mciteBstWouldAddEndPuncttrue
\mciteSetBstMidEndSepPunct{\mcitedefaultmidpunct}
{\mcitedefaultendpunct}{\mcitedefaultseppunct}\relax
\EndOfBibitem
\bibitem[Zimmerman \latin{et~al.}(2010)Zimmerman, Zhang, and
  Musgrave]{Zimmerman2010}
Zimmerman,~P.~M., Zhang,~Z., and Musgrave,~C.~B. {Singlet fission in pentacene
  through multi-exciton quantum states.} \emph{Nature chemistry} \textbf{2010}
  \emph{2}, 648--652\relax
\mciteBstWouldAddEndPuncttrue
\mciteSetBstMidEndSepPunct{\mcitedefaultmidpunct}
{\mcitedefaultendpunct}{\mcitedefaultseppunct}\relax
\EndOfBibitem
\bibitem[Yuki and Takeshi(2014)Yuki, and Takeshi]{Kurashige2014a}
Yuki,~K., and Takeshi,~Y. Theoretical Study of the π → π* Excited States of
  Oligoacenes: A Full π-Valence DMRG-CASPT2 Study. \emph{Bulletin of the
  Chemical Society of Japan} \textbf{2014} \emph{87}, 1071--1073\relax
\mciteBstWouldAddEndPuncttrue
\mciteSetBstMidEndSepPunct{\mcitedefaultmidpunct}
{\mcitedefaultendpunct}{\mcitedefaultseppunct}\relax
\EndOfBibitem
\bibitem[Coto \latin{et~al.}(2015)Coto, Sharifzadeh, Neaton, and
  Thoss]{Coto2015}
Coto,~P.~B., Sharifzadeh,~S., Neaton,~J.~B., and Thoss,~M. {Low-lying
  electronic excited states of pentacene oligomers: A comparative electronic
  structure study in the context of singlet fission}. \emph{Journal of Chemical
  Theory and Computation} \textbf{2015} \emph{11}, 147--156\relax
\mciteBstWouldAddEndPuncttrue
\mciteSetBstMidEndSepPunct{\mcitedefaultmidpunct}
{\mcitedefaultendpunct}{\mcitedefaultseppunct}\relax
\EndOfBibitem
\bibitem[Yang \latin{et~al.}(2016)Yang, Davidson, and Yang]{Yang2016}
Yang,~Y., Davidson,~E.~R., and Yang,~W. {Nature of ground and electronic
  excited states of higher acenes}. \emph{Proceedings of the National Academy
  of Sciences} \textbf{2016} 201606021\relax
\mciteBstWouldAddEndPuncttrue
\mciteSetBstMidEndSepPunct{\mcitedefaultmidpunct}
{\mcitedefaultendpunct}{\mcitedefaultseppunct}\relax
\EndOfBibitem
\bibitem[Collman \latin{et~al.}(1975)Collman, Hoard, Kim, Lang, and
  Reedzd]{Kim1975}
Collman,~J.~P., Hoard,~J.~L., Kim,~N., Lang,~G., and Reedzd,~C.~A. {Synthesis,
  Stereochemistry, and Structure-Related Properties of alpha beta gamma delta
  Tetraphenylporphinatoiron(II)}. \emph{Journal of the American Chemical
  Society} \textbf{1975} \emph{97}, 2676--2681\relax
\mciteBstWouldAddEndPuncttrue
\mciteSetBstMidEndSepPunct{\mcitedefaultmidpunct}
{\mcitedefaultendpunct}{\mcitedefaultseppunct}\relax
\EndOfBibitem
\bibitem[Goff \latin{et~al.}(1977)Goff, {La Mar}, and Reed]{Goff1977a}
Goff,~H., {La Mar},~G.~N., and Reed,~C.~A. {Nuclear Magnetic Resonance
  Investigation of Magnetic and Electronic Properties of "Intermediate Spin"
  Ferrous Porphyrin Complexes.} \emph{J. Am. Chem. Soc.} \textbf{1977}
  \emph{99}, 3641--3646\relax
\mciteBstWouldAddEndPuncttrue
\mciteSetBstMidEndSepPunct{\mcitedefaultmidpunct}
{\mcitedefaultendpunct}{\mcitedefaultseppunct}\relax
\EndOfBibitem
\bibitem[Kitagawa and Teraoaka(1979)Kitagawa, and Teraoaka]{Walker2013}
Kitagawa,~T., and Teraoaka,~J. {The Resonance Raman Spectra of
  Intermediate-spin Ferrous Porphyrin}. \emph{Chemical Physics Letters}
  \textbf{1979} \emph{63}, 443--446\relax
\mciteBstWouldAddEndPuncttrue
\mciteSetBstMidEndSepPunct{\mcitedefaultmidpunct}
{\mcitedefaultendpunct}{\mcitedefaultseppunct}\relax
\EndOfBibitem
\bibitem[Dolphin \latin{et~al.}(1976)Dolphin, Sams, Tsin, and
  Wong]{Dolphin1976}
Dolphin,~D., Sams,~J.~R., Tsin,~T.~B., and Wong,~K.~L. {Synthesis and Mossbauer
  spectra of octaethylporphyrin ferrous complexes.} \emph{Journal of the
  American Chemical Society} \textbf{1976} \emph{98}, 6970--5\relax
\mciteBstWouldAddEndPuncttrue
\mciteSetBstMidEndSepPunct{\mcitedefaultmidpunct}
{\mcitedefaultendpunct}{\mcitedefaultseppunct}\relax
\EndOfBibitem
\bibitem[Lang \latin{et~al.}(1978)Lang, Spartalian, Reed, and
  Collman]{Lang1978}
Lang,~G., Spartalian,~K., Reed,~C.~A., and Collman,~J.~P. {M{\"{o}}ssbauer
  effect study of the magnetic properties of
  {\textless}i{\textgreater}S{\textless}/i{\textgreater} =1 ferrous
  tetraphenylporphyrin}. \emph{The Journal of Chemical Physics} \textbf{1978}
  \emph{69}, 5424--5427\relax
\mciteBstWouldAddEndPuncttrue
\mciteSetBstMidEndSepPunct{\mcitedefaultmidpunct}
{\mcitedefaultendpunct}{\mcitedefaultseppunct}\relax
\EndOfBibitem
\bibitem[Boyd \latin{et~al.}(1979)Boyd, Buckingham, McMeeking, and
  Mitra]{Boyd1979}
Boyd,~P. D.~W., Buckingham,~D.~A., McMeeking,~R.~F., and Mitra,~S.
  {Paramagnetic Anisotropy, Average Magnetic Susceptibility, and Electronic
  Structure of Intermediate-Spin S = 1
  (5,10,15,20-Tetraphenylporphyrin)iron(II)}. \emph{Inorganic Chemistry}
  \textbf{1979} \emph{18}\relax
\mciteBstWouldAddEndPuncttrue
\mciteSetBstMidEndSepPunct{\mcitedefaultmidpunct}
{\mcitedefaultendpunct}{\mcitedefaultseppunct}\relax
\EndOfBibitem
\bibitem[Mispelter \latin{et~al.}(1980)Mispelter, Momenteau, and
  Lhoste]{Mispelter1980}
Mispelter,~J., Momenteau,~M., and Lhoste,~J.~M. {Proton magnetic resonance
  characterization of the intermediate (
  {\textless}i{\textgreater}S{\textless}/i{\textgreater} =1) spin state of
  ferrous porphyrins}. \emph{The Journal of Chemical Physics} \textbf{1980}
  \emph{72}, 1003--1012\relax
\mciteBstWouldAddEndPuncttrue
\mciteSetBstMidEndSepPunct{\mcitedefaultmidpunct}
{\mcitedefaultendpunct}{\mcitedefaultseppunct}\relax
\EndOfBibitem
\bibitem[Strauss \latin{et~al.}(1985)Strauss, Silver, Long, Thompson, Hudgens,
  Spartalian, and Iberslb]{Straws1985}
Strauss,~S.~H., Silver,~M.~E., Long,~K.~M., Thompson,~R.~C., Hudgens,~R.~A.,
  Spartalian,~K., and Iberslb,~J.~A. {Comparison of the Molecular and
  Electronic Structures of (2,3,7,8,12,13,17,18-Octaethylporphyrinato)iron( 11)
  and 4207 porphyrinato)iron( 11) Steven}. \emph{Journal of the American
  Chemical Society} \textbf{1985} \emph{107}, 4207--4215\relax
\mciteBstWouldAddEndPuncttrue
\mciteSetBstMidEndSepPunct{\mcitedefaultmidpunct}
{\mcitedefaultendpunct}{\mcitedefaultseppunct}\relax
\EndOfBibitem
\bibitem[Choe \latin{et~al.}(1998)Choe, Hashimoto, Nakano, and Hirao]{Choe1998}
Choe,~Y.-K., Hashimoto,~T., Nakano,~H., and Hirao,~K. {Theoretical study of the
  electronic ground state of iron (II) porphine}. \emph{Chem. Phys. Lett.}
  \textbf{1998} \emph{295}, 380--388\relax
\mciteBstWouldAddEndPuncttrue
\mciteSetBstMidEndSepPunct{\mcitedefaultmidpunct}
{\mcitedefaultendpunct}{\mcitedefaultseppunct}\relax
\EndOfBibitem
\bibitem[Choe \latin{et~al.}(1999)Choe, Nakajima, Hirao, and Lindh]{Choe1999}
Choe,~Y.-k., Nakajima,~T., Hirao,~K., and Lindh,~R. {Theoretical study of the
  electronic ground state of iron " II {\ldots} porphine . II}. \emph{Journal
  of Chemical Physics} \textbf{1999} \emph{111}, 3837--3845\relax
\mciteBstWouldAddEndPuncttrue
\mciteSetBstMidEndSepPunct{\mcitedefaultmidpunct}
{\mcitedefaultendpunct}{\mcitedefaultseppunct}\relax
\EndOfBibitem
\bibitem[Pierloot(2003)]{Pierloot2009}
Pierloot,~K. {The CASPT2 method in inorganic electronic spectroscopy: from
  ionic transition metal to covalent actinide complexes}. \emph{Molecular
  Physics} \textbf{2003} \emph{101}, 2083 -- 2094\relax
\mciteBstWouldAddEndPuncttrue
\mciteSetBstMidEndSepPunct{\mcitedefaultmidpunct}
{\mcitedefaultendpunct}{\mcitedefaultseppunct}\relax
\EndOfBibitem
\bibitem[Radoń and Pierloot(2008)Radoń, and Pierloot]{Radon2008}
Radoń,~M., and Pierloot,~K. {Binding of CO, NO, and O
  {\textless}sub{\textgreater}2{\textless}/sub{\textgreater} to Heme by Density
  Functional and Multireference ab Initio Calculations}. \emph{The Journal of
  Physical Chemistry A} \textbf{2008} \emph{112}, 11824--11832\relax
\mciteBstWouldAddEndPuncttrue
\mciteSetBstMidEndSepPunct{\mcitedefaultmidpunct}
{\mcitedefaultendpunct}{\mcitedefaultseppunct}\relax
\EndOfBibitem
\bibitem[Vancoillie \latin{et~al.}(2011)Vancoillie, Zhao, Tran, Hendrickx, and
  Pierloot]{Vancoillie2011}
Vancoillie,~S., Zhao,~H., Tran,~V.~T., Hendrickx,~M. F.~A., and Pierloot,~K.
  {Multiconfigurational second-order perturbation theory restricted active
  space (RASPT2) studies on mononuclear first-row transition-metal systems}.
  \emph{Journal of Chemical Theory and Computation} \textbf{2011} \emph{7},
  3961--3977\relax
\mciteBstWouldAddEndPuncttrue
\mciteSetBstMidEndSepPunct{\mcitedefaultmidpunct}
{\mcitedefaultendpunct}{\mcitedefaultseppunct}\relax
\EndOfBibitem
\bibitem[Phung \latin{et~al.}(2016)Phung, Wouters, and Pierloot]{Phung2016}
Phung,~Q.~M., Wouters,~S., and Pierloot,~K. {Cumulant Approximated Second-Order
  Perturbation Theory Based on the Density Matrix Renormalization Group for
  Transition Metal Complexes: A Benchmark Study}. \emph{Journal of Chemical
  Theory and Computation} \textbf{2016} \emph{12}, 4352--4361\relax
\mciteBstWouldAddEndPuncttrue
\mciteSetBstMidEndSepPunct{\mcitedefaultmidpunct}
{\mcitedefaultendpunct}{\mcitedefaultseppunct}\relax
\EndOfBibitem
\bibitem[Sun \latin{et~al.}(2017)Sun, Yang, and Chan]{Sun}
Sun,~Q., Yang,~J., and Chan,~G. K.~L. {A general second order complete active
  space self-consistent-field solver for large-scale systems}. \emph{Chemical
  Physics Letters} \textbf{2017} \emph{683}, 291--299\relax
\mciteBstWouldAddEndPuncttrue
\mciteSetBstMidEndSepPunct{\mcitedefaultmidpunct}
{\mcitedefaultendpunct}{\mcitedefaultseppunct}\relax
\EndOfBibitem
\bibitem[Olivares-Amaya \latin{et~al.}(2015)Olivares-Amaya, Hu, Nakatani,
  Sharma, Yang, and Chan]{Olivares-Amaya2015}
Olivares-Amaya,~R., Hu,~W., Nakatani,~N., Sharma,~S., Yang,~J., and Chan,~G.
  K.-L. {The ab-initio density matrix renormalization group in practice.}
  \emph{J. Chem. Phys.} \textbf{2015} \emph{142}, 034102\relax
\mciteBstWouldAddEndPuncttrue
\mciteSetBstMidEndSepPunct{\mcitedefaultmidpunct}
{\mcitedefaultendpunct}{\mcitedefaultseppunct}\relax
\EndOfBibitem
\bibitem[{Holmes} \latin{et~al.}(2017){Holmes}, {Umrigar}, and
  {Sharma}]{2017arXiv170803456H}
{Holmes},~A.~A., {Umrigar},~C.~J., and {Sharma},~S. {Excited states using
  semistochastic heat-bath configuration interaction}. \emph{ArXiv e-prints}
  \textbf{2017} \relax
\mciteBstWouldAddEndPunctfalse
\mciteSetBstMidEndSepPunct{\mcitedefaultmidpunct}
{}{\mcitedefaultseppunct}\relax
\EndOfBibitem
\bibitem[Groenhof \latin{et~al.}(2005)Groenhof, Swart, Ehlers, and
  Lammertsma]{Groenhof2005}
Groenhof,~A.~R., Swart,~M., Ehlers,~A.~W., and Lammertsma,~K. {Electronic
  ground states of iron porphyrin and of the first species in the catalytic
  reaction cycle of cytochrome P450s}. \emph{Journal of Physical Chemistry A}
  \textbf{2005} \emph{109}, 3411--3417\relax
\mciteBstWouldAddEndPuncttrue
\mciteSetBstMidEndSepPunct{\mcitedefaultmidpunct}
{\mcitedefaultendpunct}{\mcitedefaultseppunct}\relax
\EndOfBibitem
\bibitem[Sharma and Alavi(2015)Sharma, and Alavi]{Sharma2015}
Sharma,~S., and Alavi,~A. {Multireference linearized coupled cluster theory for
  strongly correlated systems using matrix product states}. \emph{The Journal
  of Chemical Physics} \textbf{2015} \emph{143}, 102815\relax
\mciteBstWouldAddEndPuncttrue
\mciteSetBstMidEndSepPunct{\mcitedefaultmidpunct}
{\mcitedefaultendpunct}{\mcitedefaultseppunct}\relax
\EndOfBibitem
\bibitem[Sharma \latin{et~al.}(2016)Sharma, Jeanmairet, and Alavi]{Sharma2016d}
Sharma,~S., Jeanmairet,~G., and Alavi,~A. {Quasi-degenerate perturbation theory
  using matrix product states.} \emph{The Journal of chemical physics}
  \textbf{2016} \emph{144}, 034103\relax
\mciteBstWouldAddEndPuncttrue
\mciteSetBstMidEndSepPunct{\mcitedefaultmidpunct}
{\mcitedefaultendpunct}{\mcitedefaultseppunct}\relax
\EndOfBibitem
\bibitem[Jeanmairet \latin{et~al.}(2017)Jeanmairet, Sharma, Alavi, Jeanmairet,
  Sharma, and Alavi]{Jeanmairet2017}
Jeanmairet,~G., Sharma,~S., Alavi,~A., Jeanmairet,~G., Sharma,~S., and
  Alavi,~A. {Stochastic multi-reference perturbation theory with application to
  the linearized coupled cluster method Stochastic multi-reference perturbation
  theory with application to the linearized coupled cluster method}.
  \textbf{2017} \emph{044107}\relax
\mciteBstWouldAddEndPuncttrue
\mciteSetBstMidEndSepPunct{\mcitedefaultmidpunct}
{\mcitedefaultendpunct}{\mcitedefaultseppunct}\relax
\EndOfBibitem
\bibitem[Fink(2006)]{Fink2006}
Fink,~R.~F. {Two new unitary-invariant and size-consistent perturbation
  theoretical approaches to the electron correlation energy}. \emph{Chem. Phys.
  Lett.} \textbf{2006} \emph{428}, 461--466\relax
\mciteBstWouldAddEndPuncttrue
\mciteSetBstMidEndSepPunct{\mcitedefaultmidpunct}
{\mcitedefaultendpunct}{\mcitedefaultseppunct}\relax
\EndOfBibitem
\bibitem[Fink(2009)]{Fink2009}
Fink,~R.~F. {The multi-reference retaining the excitation degree perturbation
  theory: A size-consistent, unitary invariant, and rapidly convergent
  wavefunction based ab initio approach}. \emph{Chemical Physics} \textbf{2009}
  \emph{356}, 39--46\relax
\mciteBstWouldAddEndPuncttrue
\mciteSetBstMidEndSepPunct{\mcitedefaultmidpunct}
{\mcitedefaultendpunct}{\mcitedefaultseppunct}\relax
\EndOfBibitem
\end{mcitethebibliography}

\end{document}